\documentclass[%
 reprint,
 amsmath,amssymb,
 aps,
prc,
]{revtex4-2}

\usepackage[caption=false]{subfig}
\usepackage{soul}
\usepackage{graphicx}
\usepackage{dcolumn}
\usepackage{bm}

\usepackage[bottom]{footmisc}
\usepackage{slashed}
\usepackage{hyperref}
\usepackage{float}
\usepackage{soul}

\usepackage{ulem}
\usepackage{color}

\begin{document}

\preprint{APS/123-QED}

\title{Neutron star matter based on a parity doublet model \\including the $a_0$(980) meson}

\author{Yuk Kei Kong}
\email{yukkekong2-c@hken.phys.nagoya-u.ac.jp}
\affiliation{Department of Physics, Nagoya University, Nagoya 464-8602, Japan}%
\author{Takuya Minamikawa}
\email{minamikawa@hken.phys.nagoya-u.ac.jp}
\affiliation{Department of Physics, Nagoya University, Nagoya 464-8602, Japan}%

\author{Masayasu Harada}
\email{harada@hken.phys.nagoya-u.ac.jp}
\affiliation{Department of Physics, Nagoya University, Nagoya 464-8602, Japan}
\affiliation{Kobayashi-Maskawa Institute for the Origin of Particles and the Universe, Nagoya University, Nagoya, 464-8602, Japan}
\affiliation{Advanced Science Research Center, Japan Atomic Energy Agency, Tokai 319-1195, Japan}




\date{\today}

\begin{abstract}
We study the effect of the isovector-scalar meson $a_0$(980) on the properties of nuclear matter and  the neutron star (NS) matter by constructing a parity doublet model with including the $a_0$ meson based on the chiral SU(2)$_L\times$SU(2)$_R$ symmetry.
We also include the $\omega$-$\rho$ mixing contribution to adjust the slope parameter at the saturation.
We find that, when the chiral invariant mass of nucleon $m_0$ is smaller than about $800$\,MeV, the existence of $a_0$(980) enlarges the symmetry energy by strengthening the repulsive $\rho$ meson coupling. On the other hand, for large $m_0$ where the Yukawa coupling of $a_0$(980) to nucleon is small, the symmetry energy is reduced by the effect of $\omega$-$\rho$ mixing.
We then construct the equation of state (EoS) of a neutron star matter to obtain the mass-radius  relation of NS. 
We find that, in most choices of $m_0$, the existence of $a_0$(980) stiffens the EoS and makes the radius of NS larger. 
We then constrain the chiral invariant mass of nucleon from the observational data of NS, and find that $580 \,\text{ MeV} \lesssim m_0 \lesssim 860 \,\text{ MeV} $ for $L_0=57.7$ MeV.
\end{abstract}

\maketitle


\section{Introduction}\label{section:1}


Spontaneous chiral symmetry breaking is one of the most important properties in the low-energy hadron physics, which is expected to generate a part of hadron masses and cause the mass difference between chiral partners.
In particular, it is interesting to study the origin of nucleon mass in terms of the chiral symmetry structure.

In the traditional linear sigma model, the entire nucleon mass is generated from the spontaneous chiral symmetry breaking, in which the chiral partner to ordinary nucleon is the nucleon itself.
In the parity doublet model (PDM) proposed in Ref.~\cite{PhysRevD.39.2805}, on the other hand, an excited nucleon such as $N(1535)$ is regarded as the chiral partner to ordinary nucleon. 
The spontaneous symmetry breaking generates the mass difference between them. 
If the chiral symmetry was not broken, their masses would be degenerated into, so called, the chiral invariant mass $m_0$.
The existence of the chiral invariant mass and its possible relation to the parity doubling structure is also supported by the lattice QCD simulation~\cite{PhysRevD.92.014503,2017JHEP...06..034A}.
In addition, recent analysis based on the QCD sum rules~\cite{PhysRevD.105.014014} also supports the existence which suggest that the origin of the chiral invariant mass is the gluon condensate. 
Therefore, quantitative and qualitative study of the chiral invariant mass will help us to understand the origin of hadron masses.

There are several analyses to determine the value of $m_0$ by studying the nucleon properties at vacuum.
For example, the analysis in Ref.~\cite{10.1143/PTP.106.873} shows that $m_0$ is smaller than $500$\,MeV using the decay width of $N(1535)$, while Ref.~\cite{Yamazaki:2018stk} includes higher derivative interaction which makes the large $m_0$ consistent with the decay width. 

The chiral symmetry is expected to be partially restored in the high density region, study of which will provide some information on the chiral invariant mass.
Actually, the PDM is applied to study the high density matter in several analyses such as in 
Refs.~\cite{Hatsuda:1988mv, Zschiesche:2006zj, Dexheimer:2007tn, Dexheimer:2008cv, Sasaki:2010bp, Sasaki:2011ff,%
Gallas:2011qp, Paeng:2011hy,%
Steinheimer:2011ea,Dexheimer:2012eu, Paeng:2013xya,Benic:2015pia,Motohiro:2015taa,%
Mukherjee:2016nhb,Suenaga:2017wbb,Takeda:2017mrm,Mukherjee:2017jzi,Paeng:2017qvp,%
Marczenko:2017huu,Abuki:2018ijb,Marczenko:2018jui,Marczenko:2019trv,Yamazaki:2019tuo,%
Harada:2019oaq,Marczenko:2020jma,Harada:2020etl,%
PhysRevC.103.045205,Marczenko:2021uaj,PhysRevC.104.065201,Marczenko:2022hyt,%
PhysRevC.106.065205,Minamikawa:2023eky,Marczenko:2023ohi%
}.
Recently in Refs.~\cite{PhysRevC.103.045205,PhysRevC.104.065201,PhysRevC.106.065205,Minamikawa:2023eky}, the EoS of neutron star (NS) matter constructed from an extended PDM ~\cite{Motohiro:2015taa} is connected to the one from
the NJL-type quark model following Refs.~\cite{Baym_2018,Baym_2019}.
The analysis of Ref.~\cite{PhysRevC.103.045205} use the observational data of NS given in 
Refs.~\cite{NANOGrav:2019jur,LIGOScientific:2017vwq,LIGOScientific:2017ync, TheLIGOScientific:2017qsa,LIGOScientific:2018cki,Miller:2019cac,Riley:2019yda}
to put constraint to the chiral invariant mass $m_0$ as $600$\,MeV$\lesssim m_0 \lesssim 900$\,MeV, which was updated in Refs.~\cite{PhysRevC.106.065205,Minamikawa:2023eky} to $400$\,MeV$\lesssim m_0 \lesssim 700$\,MeV by considering the effect of anomaly as well as new data 
analysis~\cite{Fonseca:2021wxt,De:2018uhw,Radice:2017lry}.

In recent decades, increasing attention is paid to the effect of isovector-scalar $a_0$(980) meson (or called $\delta$ meson) on asymmetric matter such as NS because it is expected to provide a different effect to neutrons and protons. References.~\cite{Kubis_1997,https://doi.org/10.48550/arxiv.astro-ph/9802303,Miyatsu_2022,Li_2022,https://doi.org/10.48550/arxiv.2209.02861,Thakur_2022,Liu_2005,PhysRevC.80.025806,Gaitanos_2004,PhysRevC.67.015203,PhysRevC.65.045201} use Walecka-type relativistic mean-field (RMF) models, and Refs.~\cite{PhysRevC.90.055801,PhysRevC.84.054309} use density-dependent RMF models to study the effect of $a_0$(980) meson to the symmetry energy as well as to the EoS of asymmetric matter.
It was pointed that the existence of $a_0$ meson increases the symmetry energy~\cite{Kubis_1997,Miyatsu_2022,Li_2022,Liu_2005,PhysRevC.80.025806,Gaitanos_2004,PhysRevC.67.015203,PhysRevC.65.045201},
and that it stiffens the NS EoS~\cite{https://doi.org/10.48550/arxiv.astro-ph/9802303,Liu_2005,Li_2022,Miyatsu_2022,Thakur_2022} and asymmetric matter EoS~\cite{PhysRevC.84.054309}.
Since there is no study on the effect of $a_0$ meson to the matter properties in PDM to our best knowledge, we study the effect of $a_0$ meson in the PDM to the symmetry energy and NS properties.

In this work, we construct a PDM with including the $a_0(980)$ meson based on the chiral SU(2)$_L\times$SU(2)$_R$ symmetry.
We first study the effect of $a_0$ meson to the symmetry energy for several choices of given values of the chiral invariant mass $m_0$.  We will show that the symmetry energy is enhanced for most choices of $m_0$, but is reduced for $m_0=900$\,MeV.
We then obtain the equation of state (EoS) of neutron star (NS) matter to compute the $M$-$R$ relation of NS. 
We think that it is not suitable to use the present hadronic model in the high density region of the NS since the model includes only nucleon and its chiral partner for baryons, while it is expected that the strangeness may appear in the high density region.
Thus, in the present analysis, we adopt a crossover prescription proposed in Ref.~\cite{Baym_2018} to construct a unified EoS
by interpolating an NJL-type quark model with the EoS from the PDM
as done in Ref.~\cite{PhysRevC.103.045205}.
By comparing the resultant $M$-$R$ relation to the observational data of NS, we obtain the constraint to the chiral invariant mass as $580 \,\text{ MeV} \lesssim m_0 \lesssim 860 \,\text{ MeV} $ for $L_0=57.7$ MeV.






This work is organized as follows. We describe our parity doublet model in Section~\ref{section:2}. Then, we construct the hadronic EoS under the mean-field approximation in Section~\ref{section:3}. 
In Section~\ref{section:4}, we compute the symmetry energy and study the effect of $a_0$(980) on it. 
In Section~\ref{section:5}, we construct the neutron star EoS by interpolating the EoS for hadronic matter with the EoS of quark matter from the NJL-type quark model, 
and compute the $M$-$R$ relation.
We then study the impact of $a_0(980)$ meson on neutron star EoS as well as the $M$-$R$ relation. Furthermore, we compare the results to the observational data of the neutron star and give a constraint to the chiral invariant mass of the nucleon. Finally, this work is summarized in Section~\ref{section:6}.

\section{Formalism}\label{section:2}


\subsection{Parity doublet model with iso-vector scalar meson}

To study the $a_0$(980) effect, we construct an SU(2)$_L \times $SU(2)$_R$ parity doublet model. The effective Lagrangian is given by \begin{equation}
    \mathcal{L} = \mathcal{L}_N + \mathcal{L}_M  +  \mathcal{L}_{V}\ ,
\end{equation} where the Lagrangian is separated into three parts: the nucleon Lagrangian $\mathcal{L}_N$, the scalar meson Lagrangian $\mathcal{L}_M$, and the vector meson Lagrangian $\mathcal{L}_{V}$. 
In ${\mathcal L}_M$,
the scalar meson field $M$ is introduced as the $(2,2)_{-2}$ representation under the SU(2)$_L \times $SU(2)$_R \times $U(1)$_A$ symmetry which transforms as 
\begin{equation}
    M \rightarrow e^{-2i\theta_A} g_L M g_R^{\dagger}\ ,
\end{equation} 
where $g_{R,L} \in \mbox{SU}(2)_{R,L} $ 
and $e^{-2i\theta_A} \in\mbox{U}(1)_A$.
We parameterize $M$ as 
\begin{equation}
    M = [\sigma + i\vec{\pi} \cdot \vec{\tau}] -   [ \vec{a_0} \cdot \vec{\tau} + i\eta],
\end{equation} where $\sigma, \vec{\pi}, \vec{a_0}, \eta$ are scalar meson fields, $\vec{\tau}$ are the Pauli matrices. Notice that the vacuum expectation value (VEV) of $M$ is 
\begin{equation}
\begin{aligned}
    \langle 0 | M | 0 \rangle = \begin{pmatrix}
\sigma_0 & 0\\
0 & \sigma_0
\end{pmatrix},
\end{aligned}
\end{equation} 
where $\sigma_0 = \langle 0|\sigma|0 \rangle$ is the VEV of the $\sigma$ field.
The explicit form of the Lagrangian $\mathcal{L}_M$ is given by 
\begin{equation}
\mathcal{L}_M = \frac{1}{4} \mbox{tr}\left[\partial_\mu M \partial^\mu M^\dagger \right] - V_M\ ,
\end{equation} 
where $V_M$ is the potential for $M$. 
In the present model, $V_M$ is taken as 
\begin{eqnarray}
V_M&=&-\frac{\bar{\mu}^2}{4}\mbox{tr}[M^\dagger M ]+\frac{\lambda_{41}}{8}\mbox{tr}[(M^\dagger M)^2]\nonumber \\ 
& &-\frac{\lambda_{42}}{16}\{ \mbox{tr}[M^\dagger M]\}^2-\frac{\lambda_{61}}{12}\mbox{tr}[(M^\dagger M)^3]\nonumber \\ 
 & &-\frac{\lambda_{62}}{24}\mbox{tr}[(M^\dagger M)^2]\mbox{tr}[M^\dagger M]-\frac{\lambda_{63}}{48}\{ \mbox{tr}[M^\dagger M]\}^3\nonumber \\
 & &-\frac{m^2_{\pi}f_{\pi}}{4}\mbox{tr}[M+M^\dagger]-\frac{K}{8}\{detM+detM^\dagger\} \ .
\label{VM}
\end{eqnarray} 
In the above potential, the first three terms account for the most general possible terms that are invariant under SU(2)$_L \times $SU(2)$_R \times $U(1)$_A$ symmetry with the mass dimension less than or equal to four. 
The terms with the coefficients $\lambda_{61}$, $\lambda_{62}$ and $\lambda_{63}$ are
the six-point interaction terms for $\sigma$ field that were introduced in Ref.~\cite{PhysRevC.92.025201} to reproduce the nuclear saturation properties.
We note that
all possible six-point interactions are included here.
In the large $N_c$ expansion of QCD, these terms are counted as 
$\mbox{tr}[(M^\dagger M )^3] \sim O(N_c)$, $\mbox{tr}[(M^\dagger M )^2]\mbox{tr}[M^\dagger M ]\sim O(1)$, and $\{ \mbox{tr}[M^\dagger M ] \}^3\sim O(1/N_c)$. Therefore, the latter two terms are suppressed compared with the first one.
In this work, we first study the effect of $a_0(980)$ meson with only the leading order term, and then
study the effect of these higher order interactions to the symmetry energy and neutron star properties.
The seventh term in $V_M$ is the explicit symmetry-breaking term due to the non-zero current quark masses, which explicitly breaks the SU(2)$_L \times $SU(2)$_R \times$U(1)$_A$ symmetry. 
The last term is introduced to account for the U(1)$_A$ anomaly.
Therefore, $\mathcal{L}_M$ is SU(2)$_L \times $SU(2)$_R \times $U(1)$_A$ invariant except the last two terms of $V_M$. For the vector meson, the iso-triplet $\rho$ meson and iso-singlet $\omega$ meson are included using the hidden local symmetry~\cite{Bando:1987br,Harada:2003jx}
to account for the repulsive force in the matter.

Finally, the baryonic Lagrangian $\mathcal{L}_N$ based on the parity doubling structure \cite{PhysRevD.39.2805,10.1143/PTP.106.873} is given by 
\begin{equation}
\begin{aligned}
\mathcal{L_N} {} & = \Bar{N_{1}}i \slashed{\partial}N_{1} + \Bar{N_{2}}i \slashed{\partial}N_{2}  \\
 & \quad - m_0 [\Bar{N}_{1}\gamma_5 N_{2} - \Bar{N}_{2}\gamma_5 N_{1}] \\
      & \quad - g_1 [\Bar{N}_{1l} M N_{1r} + \Bar{N}_{1r} M^{\dagger}  N_{1l}]\\
       & \quad - g_2 [\Bar{N}_{2r} M N_{2l} + \Bar{N}_{2l} M^{\dagger}  N_{2r}], \
\end{aligned}
\label{eq231}
\end{equation}
where $N_{ir}$($N_{il}$) ($i=1,2$) is the right-handed (left-handed) component of the nucleon fields $N_i$. By diagonalizing $\mathcal{L}_N$, we obtain two baryon fields $N_+$ and $N_-$ corresponding to the positive parity and negative parity nucleon fields, respectively.
Their masses at vacuum are obtained as~\cite{PhysRevD.39.2805,10.1143/PTP.106.873}
\begin{equation}
\begin{aligned}
    m^{\rm(vac)}_{\pm} = \frac{1}{2} \bigg[ \sqrt{(g_1+g_2)^2\sigma_0^2 + 4m_0^2} \pm (g_1 - g_2)\sigma_0 \bigg]\ .
\end{aligned}
\label{mvaj}
\end{equation} 
In the present work, $N_+$ and $N_-$ are identified as $N$(939) and $N(1535)$ respectively, so that $m_+$ ($m_-$) is the mass of $N(939)$ ($N(1535)$).

\section{Nuclear matter at non-zero density}\label{section:3}

\subsection{Mean field approximation}

To construct the nuclear matter from the model introduced in the previous section, we adopt the mean-field approximation following \cite{PhysRevC.103.045205,PhysRevC.106.065205}, by taking \begin{equation}
    \sigma(x) \rightarrow \sigma, \qquad \pi(x) \rightarrow 0, \qquad \eta(x) \rightarrow 0.
\end{equation} 
The $a_0$(980) mean field is assumed to have non-zero value only in the third axis of iso-spin as \begin{equation}
    a_0^i(x) \rightarrow a\,\delta_{i3} \ .  
\end{equation} 
Thus, the mean field for $M$ becomes 
\begin{equation}
\begin{aligned}
    \langle M \rangle = \begin{pmatrix}
\sigma - a  & 0\\
0 & \sigma + a
\end{pmatrix}.
\end{aligned}
\end{equation} 
Notice that the mean field of $a_0$(980) vanishes ($a=0$) in the symmetric nuclear matter as well as vacuum, due to the iso-spin invariance.

Redefining the parameters as
\begin{equation}
\begin{aligned}
{} & \bar{\mu}_{\sigma}^2 \equiv \bar{\mu}^2 + \frac{1}{2} K, \\
& \bar{\mu}^2_a \equiv \bar{\mu}^2 - \frac{1}{2} K  = \bar{\mu}_{\sigma}^2 - K, \\
& \lambda_4 \equiv \lambda_{41} - \lambda_{42}, \\
& \gamma_4 \equiv 3\lambda_{41} - \lambda_{42}, \\
& \lambda_6  \equiv \lambda_{61} + \lambda_{62} + \lambda_{63}, \\
& \lambda_6^{'}  \equiv \frac{4}{3}\lambda_{62} + 2\lambda_{63}, \
\label{eq2.23}
\end{aligned}
\end{equation} 
we write $V_M$ in terms of the meson mean fields as \begin{equation}
\begin{aligned}
V_M = &  - \frac{\bar{\mu}^2_{\sigma}}{2} \sigma^2 - \frac{\bar{\mu}^2_{a}}{2} a^2  +  \frac{\lambda_4}{4}  (\sigma^4 + a^4 )  + \frac{\gamma_4}{2} \sigma^2 a^2 \\ 
& {} - \frac{\lambda_6}{6}  (\sigma^6 +15\sigma^2a^4 + 15\sigma^4a^2 +a^6 )   \\
& {} +\lambda_6^{'}(\sigma^2a^4 + \sigma^4a^2) \\ 
&   - m^2_{\pi} f_{\pi} \sigma. \
\end{aligned}
\end{equation} 
We note that $\lambda_6^{'}$ is of sub-leading order in the large $N_c$ expansion.

In the mean-field approximation, the vector meson fields are taken as
\begin{equation}
    \omega_{\mu} (x) \rightarrow \omega \delta_{\mu 0}, \qquad   \rho^{i}_{\mu} (x) \rightarrow \rho \delta_{\mu 0} \delta_{i 3}.
\end{equation} 
The Lagrangian $\mathcal{L}_{V}$ of the vector mesons can then be written in terms of the mean fields of the vector mesons as
\begin{equation}
\begin{aligned}
\mathcal{L}_{V} =  & - g_{\omega} \sum_{\alpha j} \Bar{N}_{\alpha j} \gamma^0 \omega N_{\alpha j} - g_{\rho} \sum_{\alpha j} \Bar{N}_{\alpha j} \gamma^0  \frac{\tau_3}{2} \rho N_{\alpha j} \\
&  + \frac{1}{2} m^2_{\omega} \omega^2 + \frac{1}{2} m^2_{\rho} \rho^2 + \lambda_{\omega \rho} g_{\omega}^2 g_{\rho}^2 \omega^2 \rho^2. \
\end{aligned}
\end{equation} 
We note that the 
$\omega$-$\rho$ mixing is included in our model to reduce the slope parameter following Ref.~\cite{PhysRevC.106.065205}.


Here, we show that $\lambda_{\omega \rho} > 0$ is required in the present model.
The vector meson potential is written as
\begin{equation}
\begin{aligned}
V_{V} \equiv  -\frac{1}{2} m^2_{\omega} \omega^2 - \frac{1}{2} m^2_{\rho} \rho^2 - \lambda_{\omega \rho} g_{\omega}^2 g_{\rho}^2 \omega^2 \rho^2. \
\end{aligned}
\end{equation} 
The vacuum expectation values of the vector meson fields are chosen at the stationary point of the potential $V_{V}$.
 The stationary conditions 
\begin{equation}
\begin{aligned}
\frac{\partial V_{V}}{\partial \omega} = \omega[m^2_{\omega}  + 2\lambda_{\omega \rho} g_{\omega}^2 g_{\rho}^2  \rho^2 ] = 0 \ , \\ 
\frac{\partial V_{V}}{\partial \rho} = \rho[m^2_{\rho}  + 2\lambda_{\omega \rho} g_{\omega}^2 g_{\rho}^2  \omega^2 ] = 0 \ , 
\end{aligned}
\end{equation} 
imply that there are two stationary points: 
\begin{equation}
\begin{aligned}
(\omega^2,\rho^2) = (0,0) , (-\frac{m^2_{\rho}}{2\lambda_{\omega \rho} g_{\omega}^2 g_{\rho}^2},-\frac{m^2_{\omega}}{2\lambda_{\omega \rho} g_{\omega}^2 g_{\rho}^2})\ ,
\end{aligned}
\end{equation} 
which give the potential as
\begin{equation}
V_{V}=
\begin{cases}
      0, & \text{for}\ (\omega^2,\rho^2) = (0,0) \ , \\
      \frac{m^2_{\omega}m^2_{\rho}}{4\lambda_{\omega \rho} g_{\omega}^2 g_{\rho}^2}, & \text{for}\ (\omega^2,\rho^2) = (-\frac{m^2_{\rho}}{2\lambda_{\omega \rho} g_{\omega}^2 g_{\rho}^2},-\frac{m^2_{\omega}}{2\lambda_{\omega \rho} g_{\omega}^2 g_{\rho}^2})\ .
    \end{cases}
  \end{equation} 
In the present work, a vanishing vacuum expectation values of the vector meson fields is required at zero density due to the Lorentz-invariance of the vacuum. 
Therefore, we need $\lambda_{\omega \rho} > 0$ such that $(\omega^2,\rho^2) = (0,0)$ minimizes the effective potential $V_{V}$ at vacuum.

Now, the thermodynamic potential for the nucleons is written as \begin{equation}
\begin{aligned}
{} & \Omega_{N}   = - 2 \sum_{\alpha=\pm,  j=p,n} \int^{k_f} \frac{d^3p}{(2 \pi)^3} \bigg[ \mu^*_j - \omega_{\alpha j}  \bigg],
\end{aligned}
\label{OFG}
\end{equation} 
where 
$\alpha = \pm$ denotes the parity and $j=p,n$ the iso-spin of nucleons.
$\mu^*_j$ is the effective chemical potential given by \begin{equation}
\begin{aligned}
\mu^*_j \equiv (\mu_B - g_{\omega} \omega) + \frac{j}{2} (\mu_I - g_{\rho} \rho)\ ,
\end{aligned}
\label{mustar}
\end{equation} 
and $\omega_{\alpha j}$ is the energy of the nucleon
defined as $\omega_{\alpha j} = \sqrt{(\vec{p})^2 + (m^*_{\alpha j})^2}$ where $\vec{p}$ and $m^*_{\alpha j}$ are the momentum and the effective mass of the nucleon. The effective mass $m^*_{\alpha j}$ is given by 
\begin{equation}
\begin{aligned}
    m^*_{\alpha j} = {} & \frac{1}{2} \bigg[ \sqrt{(g_1+g_2)^2(\sigma - ja)^2 + 4m_0^2}\\ 
    & + \alpha(g_1 - g_2)(\sigma - ja) \bigg]\ .
\end{aligned}
\label{maj}
\end{equation} 
Notice that the masses of proton and neutron become non-degenerate in the asymmetric matter due to the non-zero mean field of $a_0$(980).


The entire thermodynamic potential for hadronic matter 
is expressed as 
\begin{equation}
\begin{aligned}
{} & \Omega_H   = \Omega_{N} \\
& \qquad \;\;  - \frac{\bar{\mu}^2_{\sigma}}{2} \sigma^2 - \frac{\bar{\mu}^2_{a}}{2} a^2  +  \frac{\lambda_4}{4}  (\sigma^4 + a^4 ) + \frac{\gamma_4}{2} \sigma^2 a^2  \\
& \qquad \;\; - \frac{\lambda_6}{6}  (\sigma^6 +15\sigma^2a^4 + 15\sigma^4a^2+ a^6 )   + \lambda_6^{'}(\sigma^2a^4 + \sigma^4a^2) \\ 
& \qquad \;\;  - m^2_{\pi} f_{\pi} \sigma - \frac{1}{2} m^2_{\omega} \omega^2 - \frac{1}{2} m^2_{\rho} \rho^2 - \lambda_{\omega \rho} g_{\omega}^2 g_{\rho}^2 \omega^2 \rho^2\\
& \qquad \;\; - \Omega_{0}\ ,
\end{aligned}
\label{eq36}
\end{equation} 
where we subtracted the potential at the vacuum \begin{equation}
\begin{aligned}
\Omega_0 \equiv - \frac{\bar{\mu}^2_{\sigma}}{2} f_{\pi}^2  +  \frac{\lambda_4}{4}  f_{\pi}^4  
 - \frac{\lambda_6}{6} f_{\pi}^6  - m^2_{\pi} f_{\pi}^2.
\end{aligned}
\label{eq37}
\end{equation}

\subsection{Parameters determination}

\begin{figure}[h!]
\includegraphics[scale=0.55, bb = 0 0 461 346 ]{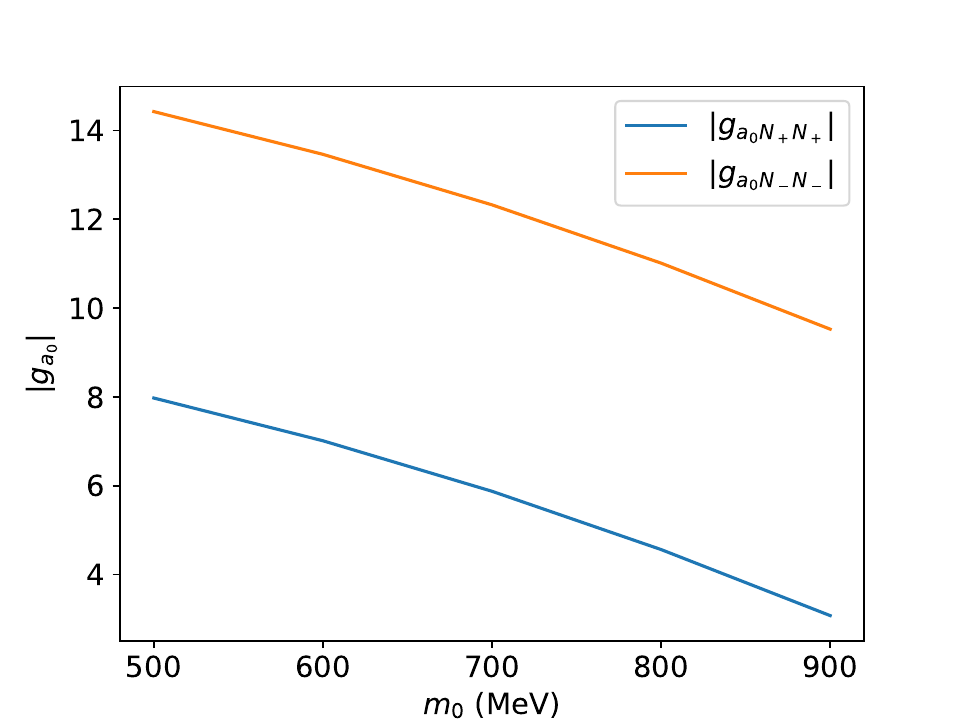}
\caption{\label{figyc}Strength of the Yukawa couplings of $a_0$ meson, $|g_{a_0 NN}|=|g_{\sigma NN}|$, as a function of the chiral invariant mass $m_0$. }
\end{figure}

In the present model, there are 12 parameters to be determined for a given value of the chiral invariant mass $m_0$: 
\begin{equation}
    g_1, g_2, \bar{\mu}^2_{\sigma}, \bar{\mu}^2_{a}, \lambda_4,\gamma_4,\lambda_6,\lambda'_6 ,g_{\omega},  g_{\rho}, \lambda_{\omega \rho}.
\end{equation} 
We determine them from the vacuum properties as well as nuclear saturation properties as follows:

The vacuum expectation value of $\sigma$ is taken to be $\sigma_0=f_{\pi}$ with the pion decay constant $f_{\pi}$=92.4 MeV. 
The Yukawa coupling constants
$g_1$ and $g_2$ are determined by fitting them to 
the nucleon masses at vacuum given by Eq.~(\ref{mvaj}), with $m_+ = m_{N}=939$\,MeV and $m_{-} = m_{N^*}=1535$\,MeV. 
The value of $g_1$, $g_2$ is given in Tables~\ref{PD1}. It is convenient to define the effective 
Yukawa couplings of $\sigma$ and $a_0(980)$ as
\begin{align}
g_{\sigma N_{\alpha j}N_{\alpha j}} & \equiv \frac{\partial m^*_{\alpha j}}{\partial \sigma} |_{\rm vacuum}\notag \\
&=  \frac{1}{2} \bigg[ \frac{(g_1+g_2)^2f_\pi}{\sqrt{(g_1+g_2)^2f_\pi^2 + 4m_0^2}} + \alpha (g_1 - g_2) \bigg]\ ,
\\
g_{a_0 N_{\alpha j}N_{\alpha j}} & \equiv \frac{\partial m^*_{\alpha j}}{\partial a} |_{\rm vacuum}= (-j)g_{\sigma N_{\alpha j}N_{\alpha j}}\ .
\label{eqyca}
\end{align}
These imply that 
$\sigma$ and $a_0$(980) meson couple to the nucleons with the same strength $|g_{\sigma NN}| = |g_{a_0 NN}|$. 
In Fig.~\ref{figyc} we show how the the Yukawa coupling of $a_0$ meson changes as $m_0$ changes, which shows that the Yukawa coupling of $a_0$ meson decreases as $m_0$ increases.
Then, the $a_0$ meson effect to the 
symmetry energy becomes smaller 
for large $m_0$ as we will show later.

Tables~\ref{PD1} also shows the parameters $\bar{\mu}^2_{\sigma}$,$\lambda_4$,$\lambda_6$, $g_{\omega}$ that are determined from the constraints at nuclear saturation.\begin{table}[h!]
\caption{\label{PD1} Values of $g_1$, $g_2$, $\bar{\mu}^2_{\sigma},  \lambda_4, \lambda_6, g_{\omega}$ for $m_0$ = $500- 900$ MeV.}
\begin{ruledtabular}
\begin{tabular}{cccccc}
\textrm{$m_0$ (MeV)}&
\textrm{500}&
\textrm{600}&
\textrm{700}&
\textrm{800}&
\textrm{900}\\
\colrule
$g_1$& 9.02 & 8.48 & 7.81 & 6.99 & 5.96 \\
$g_2$& 15.47 & 14.93 & 14.26 & 13.44 & 12.41 \\
$\bar{\mu}^2_{\sigma}/f^2_{\pi}$& 22.70   & 22.35  & 19.28    & 11.93    & 1.50 \\
{{$\lambda_4$}}&41.94   & 40.39   & 35.46    & 23.12    & 4.43 \\
{{$\lambda_6 f^2_{\pi}$}}&16.94   & 15.75   & 13.89    & 8.89     & 0.64 \\
{{$g_{\omega}$}}&11.34   & 9.13    & 7.30     & 5.66     & 3.52 \\
$\bar{\mu}^2_a/f^2_{\pi}$& $-$10.43 & $-$10.79   &   $-$13.86 &   $-$21.21  &$-$31.64 \\$\gamma_4$& 191.41 & 185.07   &   172.70 &   140.38  & 88.67 \\
\end{tabular}
\end{ruledtabular}
\end{table} The saturation properties used are summarized in Table~\ref{SP}.\begin{table}[h!]
\caption{\label{SP}Saturation properties that are used to determine the model parameters: saturation density $n_0$, binding energy $B_0$, incompressibility $K_0$, symmetry energy $S_0$, and slope parameter $L_0$.}
\begin{ruledtabular}
\begin{tabular}{ccccc}
\textrm{$n_0$[$fm^{-3}$]}&
\textrm{$B_0$[MeV]}&
\textrm{$K_0$[MeV]}&
\textrm{$S_0$[MeV]}&
\textrm{$L_0$[MeV]}\\
\colrule
0.16 & 16 & 240 & 31 & $40-80$\\
\end{tabular}
\end{ruledtabular}
\end{table}
 Then, $\bar{\mu}_{a}^2=\bar{\mu}_{\sigma}^2 - K$ is determined from $K$ given by 
 \begin{equation}
    \begin{aligned}
    K = m_{\eta}^2 - m_{\pi}^2\ . 
    \end{aligned}
\label{K} 
\end{equation} Similarly, $\gamma_4$ is given by 
\begin{equation}
    \gamma_4 = \frac{m_{a}^2 + 5 \lambda_{6} f_{\pi}^4 + \bar{\mu}^2_{a}}{f_{\pi}^2}\ ,
\end{equation} 
from the mass of $a_0$(980). 
We summarize the values of masses of mesons in Table~\ref{vacM}. 
\begin{table}[h!]
\caption{\label{vacM}
Values of meson masses at vacuum in unit of MeV.}
\begin{ruledtabular}
\begin{tabular}{ccccc}
\textrm{$\pi$}&
\textrm{$a_0$(980)}&
\textrm{$\eta$}&
\textrm{$\omega$}&
\textrm{$\rho$}\\
\colrule
140 & 980 & 550 & 783 & 776\\
\end{tabular}
\end{ruledtabular}
\end{table}

In the present analysis, 
$\lambda'_6$ is treated as a free parameter to study the high-order effect in the large $N_c$ expansion to the mesonic six-point interactions.
Since the $\lambda'_6$ term is suppressed by $1 / N_c$ compared with the $\lambda_6$ term in the large $N_c$ expansion, we assume that $|\lambda'_6| \lesssim |\lambda_6|$ is satisfied,  and take $\lambda'_6 =0, \pm \lambda_6$ to examine the effect of the higher order six-point interactions to the symmetry energy and neutron star properties. 

Tables~\ref{tablegra0} and Tables~\ref{tableLwra0} show the values of 
the parameters $g_{\rho}$ and $\lambda_{\omega \rho}$ that are determined from fitting $S_0$ and $L_0$ with $\lambda'_6 =0$.\begin{table}[h!]
\caption{\label{tablegra0}$g_{\rho}$ of model with $a_0$(980) meson and $\lambda'_6 =0$.}
\begin{ruledtabular}
\begin{tabular}{cccccc}
\textrm{$m_0$ (MeV)}&
\textrm{500}&
\textrm{600}&
\textrm{700}&
\textrm{800}&
\textrm{900}\\
\colrule
$L_0$ = 40 MeV &  19.43 & 15.52 & 13.89 & 12.64 & 11.40 \\
$L_0$ = 50 MeV& 18.75 & 15.03 & 13.35 & 12.0 & 10.69 \\
$L_0$ = 60 MeV& 18.14 & 14.59 & 12.87 & 11.45 & 10.09 \\
$L_0$ = 70 MeV& 17.58 & 14.18 & 12.44 & 10.97 & 9.59 \\
$L_0$ = 80 MeV& 17.07 & 13.81 & 12.05 & 10.54 & 9.15 \\
\end{tabular}
\end{ruledtabular}
\end{table}\begin{table}[h!]
\caption{\label{tableLwra0}$\lambda_{\omega \rho}$ of model with $a_0$(980) meson and $\lambda'_6 =0$.}
\begin{ruledtabular}
\begin{tabular}{cccccc}
\textrm{$m_0$ (MeV)}&
\textrm{500}&
\textrm{600}&
\textrm{700}&
\textrm{800}&
\textrm{900}\\
\colrule
$L_0$ = 40 MeV &  0.0127 & 0.0251 & 0.0761 & 0.2916 & 2.4595 \\
$L_0$ = 50 MeV & 0.0119 & 0.0221 & 0.0649 & 0.2415 & 1.9427 \\
$L_0$ = 60 MeV& 0.0110 & 0.0192 &  0.0537 &   0.1914 & 1.4259 \\
$L_0$ = 70 MeV& 0.0101 & 0.0162   &   0.0425 &   0.1413 & 0.9091 \\
$L_0$ = 80 MeV& 0.0092 & 0.0132 &   0.0313 &   0.0911 & 0.3923 \\ 
\end{tabular}
\end{ruledtabular}
\end{table} 
As we will discuss later, the higher-order effect is small and the values of $g_{\rho}$ and $\lambda_{\omega \rho}$ are similar for $\lambda'_6 = \pm \lambda_6$. 
We note that the inclusion of the $\omega$-$\rho$ mixing allows us to vary the slope parameter $L_0$ in the present model. Reference~\cite{universe7060182} shows that $L_0 = 57.7\pm 19$\,MeV, so we take  $L_0=40-80$ MeV as the physical input to compare the effect of $a_0$(980). 


To compare how the matter properties are affected by $a_0$(980) meson,
we eliminate the $a_0(980)$ meson by taking the mean field $a=0$ in the potential (\ref{eq36}) and the masses in Eq.~(\ref{maj}).
The parameters are determined 
by fitting them to nuclear saturation properties and vacuum properties. 
Similarly, in the model without $a_0(980)$, 
the model parameters $g_1, g_2, \bar{\mu}^2_{\sigma}, \lambda_4,\lambda_6$ are the same as in the $a_0$(980) model, since these parameters are determined from the properties of symmetric matter irrelevant for isospin asymmetry.
We notice that $\gamma_4$- and $\lambda'_6$-terms do not exist in the model without $a_0$(980) which take account of the cross interactions between $\sigma$ and $a$. 
We summarize the values of  the parameters $g_{\rho}$ and $\lambda_{\omega \rho}$ in the model without $a_0$ meson
in Tables~\ref{tablegr} and Tables~\ref{tableLwr}.
\begin{table}[h!]
\caption{\label{tablegr}$g_{\rho}$ of model without $a_0$(980) meson.}
\begin{ruledtabular}
\begin{tabular}{cccccc}
\textrm{$m_0$ (MeV)}&
\textrm{500}&
\textrm{600}&
\textrm{700}&
\textrm{800}&
\textrm{900}\\
\colrule
$L_0$ = 40 MeV &  12.48 & 10.99 & 10.72 & 10.64 & 10.61 \\
$L_0$ = 50 MeV& 10.72 & 10.01 & 9.91 & 9.90 & 9.91 \\
$L_0$ = 60 MeV& 9.54 & 9.24 & 9.25 & 9.29 & 9.34 \\
$L_0$ = 70 MeV& 8.68 & 8.63 & 8.71 & 8.78 & 8.86 \\
$L_0$ = 80 MeV& 8.02 & 8.13 & 8.26 & 8.35 & 8.44 \\
\end{tabular}
\end{ruledtabular}
\end{table}\begin{table}[h!]
\caption{\label{tableLwr}$\lambda_{\omega \rho}$ of model without $a_0$(980) meson.}
\begin{ruledtabular}
\begin{tabular}{cccccc}
\textrm{$m_0$ (MeV)}&
\textrm{500}&
\textrm{600}&
\textrm{700}&
\textrm{800}&
\textrm{900}\\
\colrule
$L_0$ = 40 MeV &  0.0554
 & 0.0857 &   0.1695 &   0.4159 & 2.5186 \\
$L_0$ = 50 MeV& 0.0451 & 0.0671 &   0.1301 &   0.3153 & 1.8903 \\
$L_0$ = 60 MeV& 0.0348 & 0.0486 &  0.0907 &   0.2147 & 1.2619 \\
$L_0$ = 70 MeV& 0.0245 & 0.0301 &   0.0513 &   0.1140 & 0.6336 \\ 
$L_0$ = 80 MeV& 0.0142 & 0.0116   &   0.0120 &   0.0134 & 0.0052 \\
\end{tabular}
\end{ruledtabular}
\end{table}

\section{effect of $a_0$(980) to the symmetry energy}\label{section:4}

In this section, we study the effect of $a_0(980)$ meson to the symmetry energy.
For a while we take $\lambda_6' = 0$, and study its effect by taking $\lambda_6' = \pm \lambda_6$ at the end of this section. 
In the present model, the symmetry energy S($n_B$) for a given density $n_B$ is expressed as 
\begin{equation}
\begin{aligned}
    S(n_B) & {} = \frac{n_B}{8} 
    \frac{\partial \mu_I}{\partial n_I} \Bigr|_{\substack{n_I=0}}\\
    & =  \frac{(k^*_+)^2}{6 \mu^*_+}   + \frac{n_B}{2}\frac{(g_{\rho}/2)^2}{m_{\rho}^2 + (2 \lambda_{\omega \rho}g_\omega^4 g_\rho^2 {{n_B^2}/{m_{\omega}^4}})} \\
    & \quad \; -  \frac{n_B}{4} \frac{m^*_+}{\mu^*_+}\frac{\partial m^*_{+n}}{\partial  n_I}\Bigr|_{\substack{n_I=0}} \ ,
    \label{SrhoBa}
\end{aligned}
\end{equation} 
where $\mu^*_+ \equiv \mu^*_p \big|_{\substack{n_I=0}}=\mu^*_n \big|_{\substack{n_I=0}}$ is the effective chemical potential for $N(939)$ in the symmetric matter, $k^*_{+} \equiv \sqrt{(\mu^*_p)^2 - (m^*_{+p})^2}  \big|_{\substack{n_I=0}} = \sqrt{(\mu^*_n)^2 - (m^*_{+n})^2}  \big|_{\substack{n_I=0}}$ the corresponding Fermi momentum, $m^*_{+} \equiv m^*_{+p} \big|_{\substack{n_I=0}} = m^*_{+n} \big|_{\substack{n_I=0}}$ the mass.
%

\begin{figure*}
    \subfloat[$L_0$ = 40 MeV\label{S40}]{%
      \includegraphics[width=0.45\textwidth]{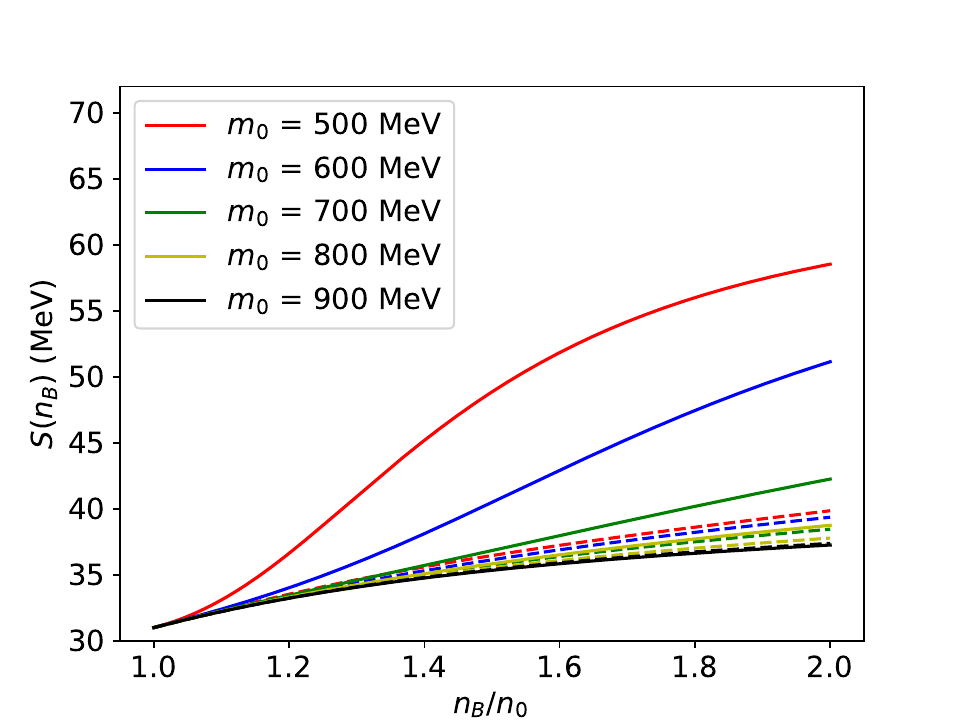}
    }
    \hfill
    \subfloat[$L_0$ = 50 MeV\label{S50}]{%
      \includegraphics[width=0.45\textwidth]{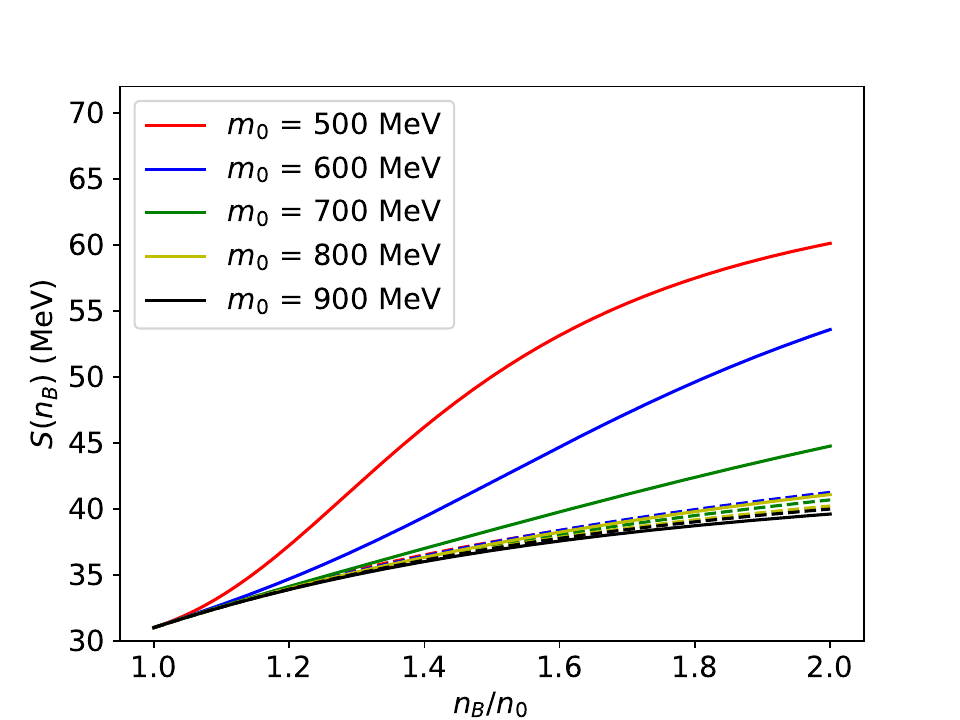}
    }
    \hfill
    \subfloat[$L_0$ = 57.7 MeV\label{S60}]{%
      \includegraphics[width=0.45\textwidth]{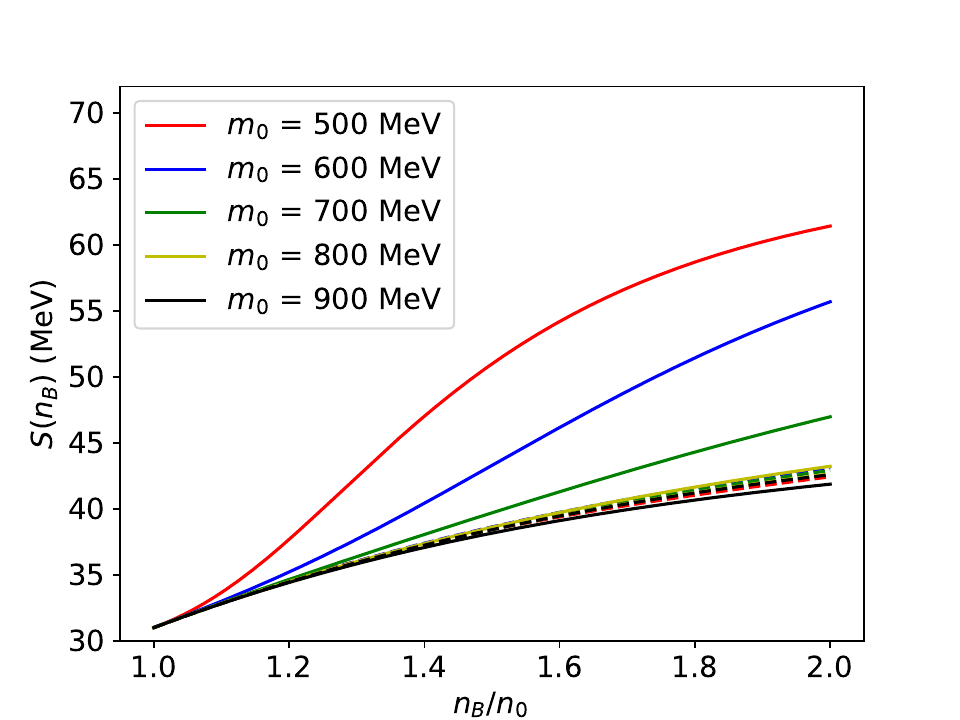}
    }
    \hfill
    \subfloat[$L_0$ = 70 MeV\label{S70}]{%
      \includegraphics[width=0.45\textwidth]{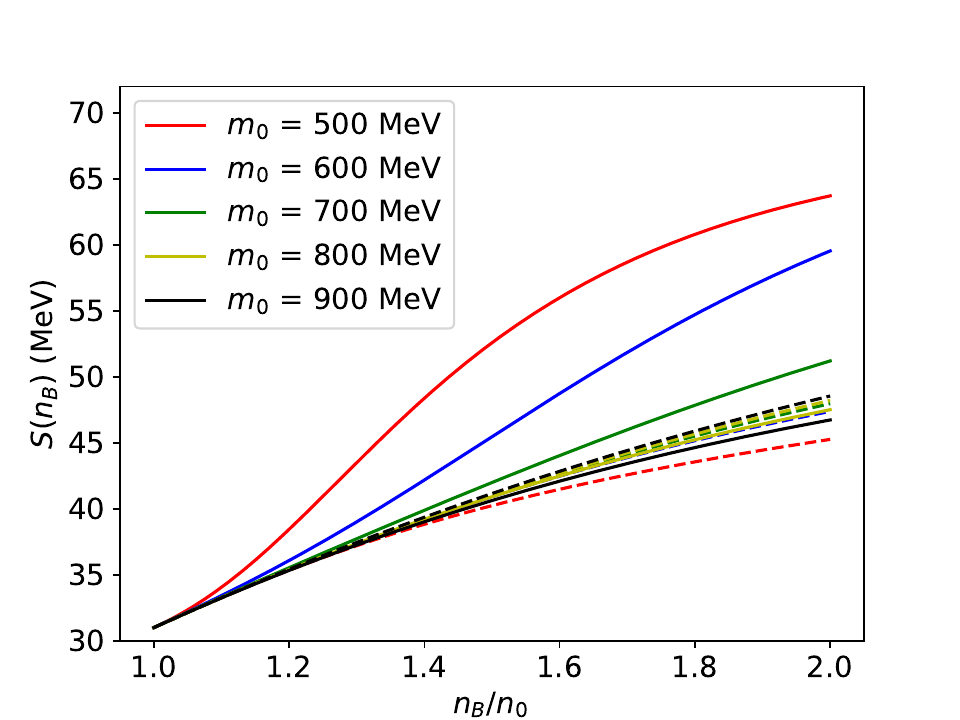}
    }
    \hfill
    \subfloat[$L_0$ = 80 MeV\label{S80}]{%
      \includegraphics[width=0.45\textwidth]{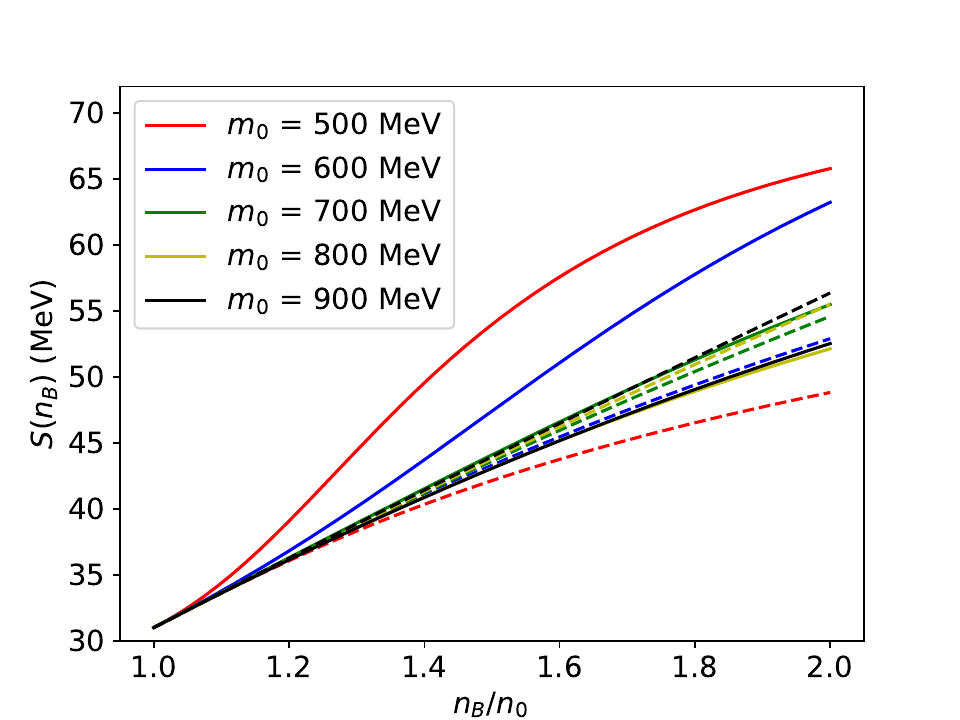}
    }
\caption{\label{symmetryenergy} (color online) Symmetry energy $S(n_B)$ for $m_0=500$-$900$\,MeV and $L_0 = 40$-$80$\,MeV. Solid curves represent the $S(n_B)$ of the model including $a_0(980)$ with $\lambda_{6}'$=0, while dashed curves show the results of the model without $a_0(980)$. }
\end{figure*}
Figure~\ref{symmetryenergy} shows the symmetry energy for $m_0=500-900 $\,MeV and $L_0=40-80 $\,MeV. 
Here we compare the results obtained with and without the existence of $a_0(980)$ meson. 
We observe that, in most cases, the symmetry energy is stiffened by the existence of $a_0$(980) and the difference of the symmetry energy between the models with same $L_0$ is larger for smaller $m_0$. 
At $n_B = 2n_0$, the symmetry energy $S(2n_0)$ is enlarged by as large as 51$\%$ in the $a_0$ model depending on the input parameters.

To further understand the nature of this stiffening of the symmetry energy in the $a_0$(980) model, we look more carefully into Eq.~(\ref{SrhoBa}) and find that there are three contributions to the symmetry energy as 
\begin{equation}
\begin{aligned}
    S(n_B)\equiv S_k(n_B) + S_{\rho}(n_B) + S_{a_0}(n_B),
    \label{Sk}
\end{aligned}
\end{equation} where $S_k(n_B)$ is the kinetic contribution of the nucleon, and $S_{\rho}(n_B)$ and $S_{a_0}(n_B)$ are the contrinutions from $\rho$ and $a_0$(980) mesons, respectively. In the following discussion, we will focus on the results for $L_0 = 57.7$\,MeV.

\begin{figure}[h!]
\includegraphics[scale=0.55]{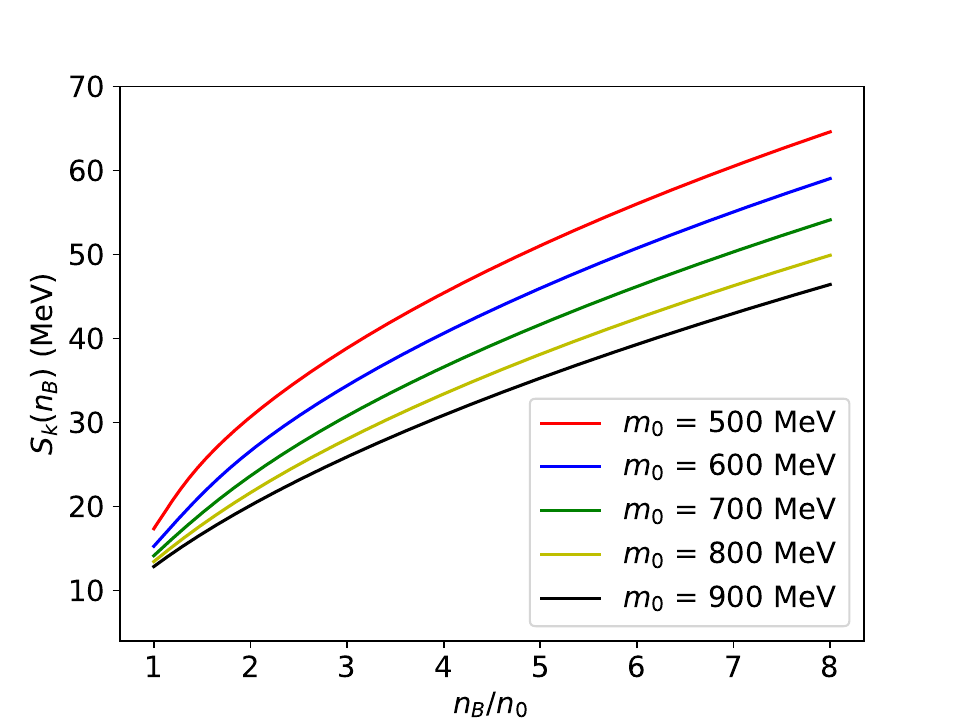}
\caption{\label{skplotmc}Density dependence of $S_{k}(n_B)$ for $m_0= 500 - 900 $\,MeV and $L_0 = 57.7 $\,MeV.
}
\label{skplotm}
\end{figure}

The kinetic contribution from the nucleons $S_k(n_B)$ is expressed as 
\begin{equation}
\begin{aligned}
    S_k(n_B)\equiv\frac{n_B}{2}\left[\frac{(k^*_+)^2}{3 \mu^*_+n_B}\right],
    \label{Sk}
\end{aligned}
\end{equation} 
which accounts for the repulsion from the extra nucleon put into the symmetric matter. The density dependence of $S_k(n_B)$ is shown in Fig.~\ref{skplotm}.
As expected, the kinetic contribution increases as the baryon number density increases. This is because higher baryon number density increases the repulsion between nucleons, and thus increases the energy of the matter.

\begin{figure}[h!]
\includegraphics[scale=0.55]{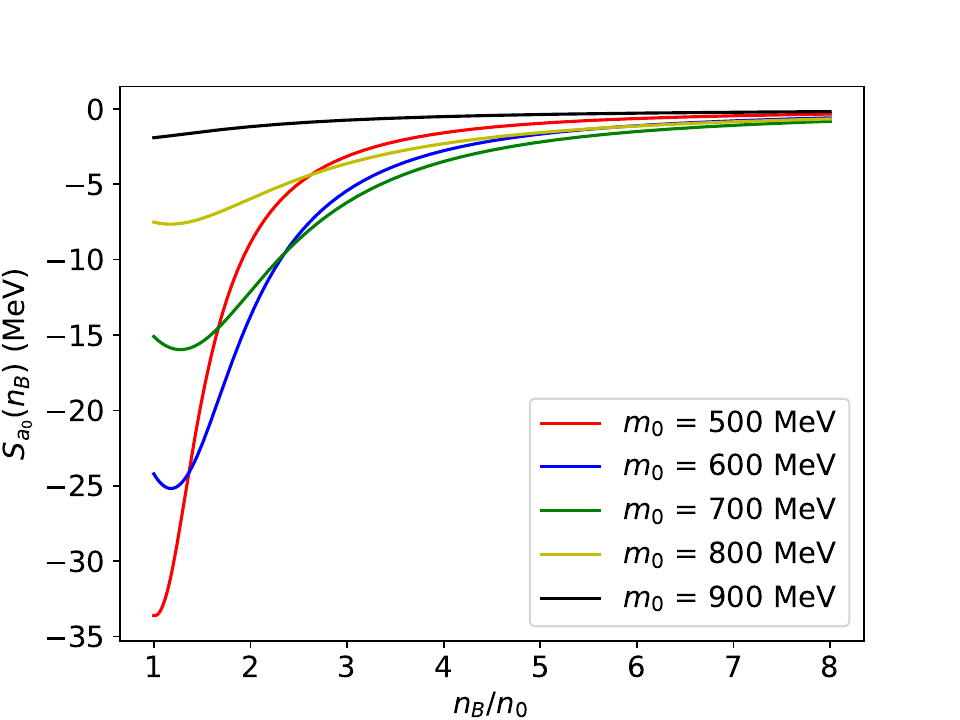}
\caption{\label{Sac}Density dependence of $S_{a_0}(n_B)$ for $m_0 = 500 - 900 $\,MeV and $L_0= 57.7 $\,MeV.
}
\label{Sa}
\end{figure}

\begin{figure}[h!]
\includegraphics[scale=0.55]{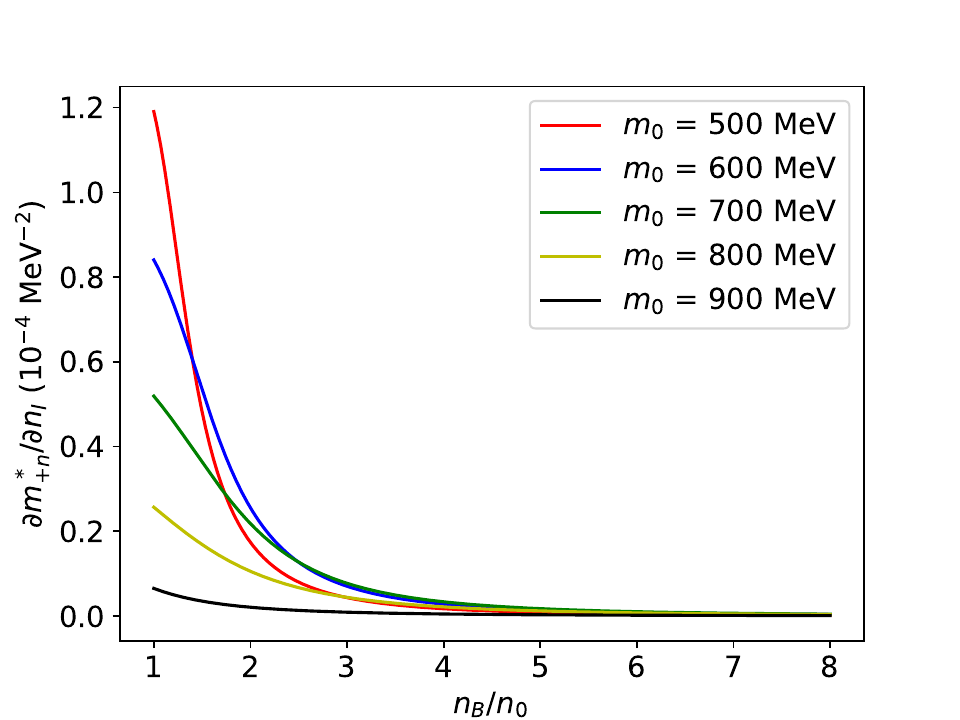}
\caption{\label{dmdrhoic}Density dependence of $\frac{\partial m^*_{+n}}{\partial  n_I}|_{n_I=0}$ for $m_0$ = 500 $-$ 900 MeV and $L_0$ = 57.7 MeV. 
}
\label{dmdrhoi}\end{figure}

The contribution from the $a_0(980)$ meson is expressed as \begin{eqnarray}
    S_{a_0}(n_B)   & {} \equiv  - \frac{n_B}{4} \frac{m^*_+}{\mu^*_+}\frac{\partial m^*_{+n}}{\partial  n_I}\Bigr|_{\substack{n_I=0}}.
    \label{F}
\end{eqnarray}  
Figure~\ref{Sa} shows the $S_{a_0}$ computed in the present model.
We notice that $S_{a_0}$ is negative and thus reduce the total symmetry energy $S(n_B)$. This is because $\frac{\partial m^*_{+n}}{\partial n_I}|_{\substack{n_I=0}}$ is always positive as shown in Fig.~\ref{dmdrhoi}.
Intuitively, this can be understood from the dependence of $m^*_{+n}$ on the mean field $a$ given in Eq.~(\ref{maj}).
If we vary the mean field $a$, $m^*_{+n}$ will also change correspondingly. However, the effective chemical potential $\mu_{n}^*$ is not depend on the mean field $a$ directly as we can see from Eq.~(\ref{mustar}). This change of the effective mass $m^*_{+n}$ due to the mean field $a$ leads to the change of the momentum of the neutron $k_{+n} = \sqrt{(\mu_{n}^*)^2 - (m^*_{+n})^2}$. When $n_I = n_p - n_n$ is increased for a fixed $n_B$, the density of the neutron $n_n$ and thus the momentum $k_{+n}$ is decreased. 
Accordingly, the effective mass of neutron is increasing as $n_I$ increase, causing a positive $\frac{\partial m_{+n}}{\partial n_I}|_{\substack{n_I=0}}$. Therefore, the $a_0$(980) meson contribution $S_{a_0}(n_B)$ reduces the total symmetry energy $S(n_B)$ in the present model. We also find that the $a_0$(980) effect on the symmetry energy is stronger for smaller $m_0$. This is because the coupling constants of $a_0(980)$ meson to the nucleon, $g_1$ and $g_2$, are larger for smaller $m_0$ as shown in Table~\ref{PD1}. 
As a result, the symmetry energy is enlarged by $a_0(980)$ meson more when $m_0$ is smaller. In addition, we notice that the $a_0(980)$ effect on the symmetry energy is decreasing as the density increases since
$\frac{\partial m_{+n}}{\partial n_I}|_{\substack{n_I=0}}$ decreases.

\begin{figure}[h!]
\includegraphics[scale=0.55]{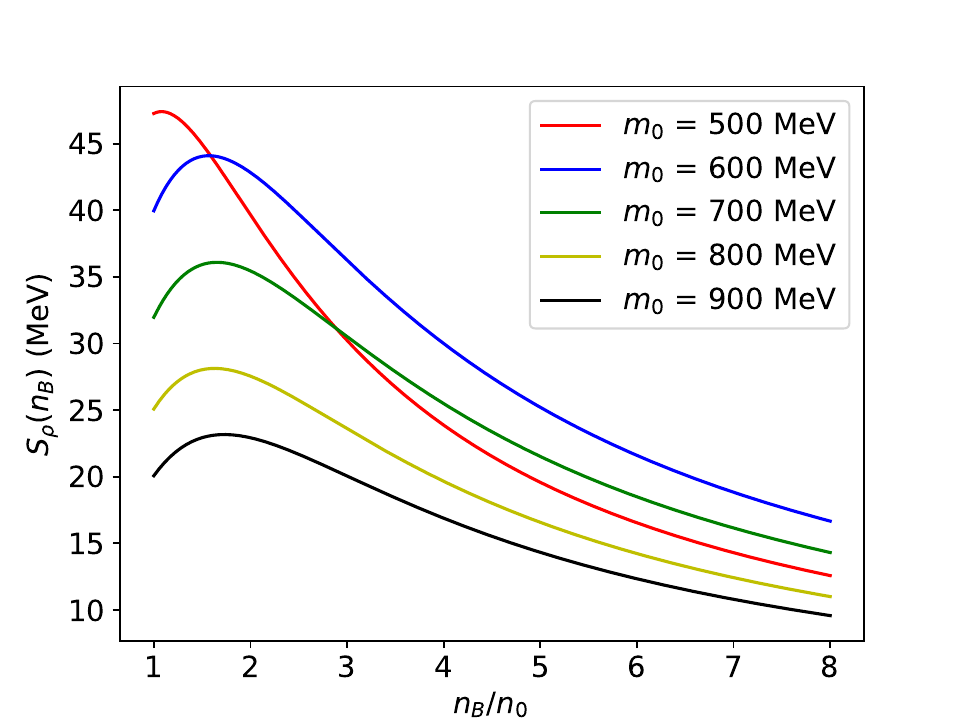}
\caption{\label{srhoplotmc}Density dependence of $S_{\rho}(n_B)$ for $m_0 = 500 - 900 $\,MeV and $L_0 = 57.7 $\,MeV.}
\label{srhoplotm}
\end{figure}

The third contribution comes from the $\rho$ meson which is expressed as
\begin{equation}
\begin{aligned}
    S_{\rho}(n_B) \equiv \frac{n_B}{2}\Big[  \frac{(g_{\rho}/2)^2}{m_{\rho}^2 + (2 \lambda_{\omega \rho}g_\omega^4 g_\rho^2 n_B^2/m_{\omega}^4)} \Big].
    \label{Srho}
\end{aligned}
\end{equation} The density dependence of $S_{\rho}$ is shown in Fig.~\ref{srhoplotm}. 
This shows that the contribution is always positive and thus provides repulsive force to the matter. 
We notice that $\omega$ meson affects the symmetry energy through $2 \lambda_{\omega \rho}g_\omega^4 g_\rho^2 {{n_B^2}/{m_{\omega}^4}}$ in the denominator.
Since $\lambda_{\omega \rho}>0$ as shown before, the $\omega$-$\rho$ mixing term always reduces the symmetry energy. In the case of large $m_0$ such as $m_0=900$ MeV where the $a_0$(980) meson effect is small, the softening effect of $\lambda_{\omega \rho}$ term overrides the stiffening effect from the $a_0$(980) meson. 
As a result, the symmetry energy $S(n_B)$ is reduced after the inclusion of $a_0$ meson.
A similar reduction of symmetry energy in the intermediate density region was also reported in  Ref.~\cite{Li_2022} which includes both the scalar meson mixing and the vector meson mixing interactions in an RMF model with the presence of isovector-scalar meson.

Moreover, we observe in Fig.~\ref{srhoplotm} that $S_{\rho}$ is increasing with increasing $n_B$ in the low-density region and becomes decreasing in the high-density region. This is understood as follows:
In the low-density region, $m_\rho^2 \gg 2 \lambda_{\omega \rho}g_\omega^4 g_\rho^2 {{n_B^2}/{m_{\omega}^4}}$ is satisfied which implies that the density dependence of $S(n_B)$ is determined by the pre-factor $n_B$.
In the high density region, on the other hand, the denominator is dominated by $2 \lambda_{\omega \rho}g_\omega^4 g_\rho^2 {{n_B^2}/{m_{\omega}^4}}$, which leads to $S_\rho(n_B) \propto 1/ n_B$.

\begin{figure}[h!]
\includegraphics[scale=0.55]{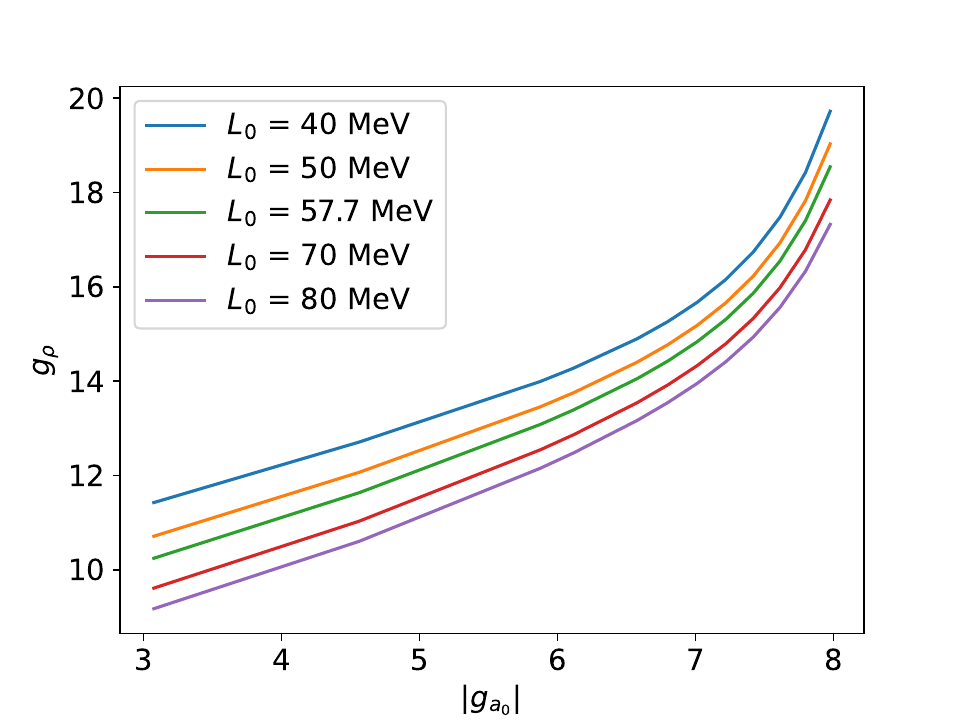}
\caption{\label{grga0}Relation between $g_{\rho}$ and $|g_{a_0 NN}|$ for $L_0 = 57.7$ MeV. 
}
\end{figure}

Based on the above properties of three contributions,
the symmetry energy can be understood as a result of the competition between the repulsive $\rho$ meson interaction (modeified by the $\omega$-$\rho$ mixing interaction) and the attractive $a_0$(980) interaction, in addition to the kinetic contribution from the nucleons. 
On the other hand, in the model without $a_0$ meson, only repulsive contributions exist. Since the symmetry energy at saturation density is fixed as $S_0=31$\,MeV in both the models with and without $a_0(980)$ meson, the $\rho$ meson coupling $g_{\rho}$ is strengthened by the existence of the attractive $a_0(980)$ contribution in the model with $a_0$ comparing to the model without $a_0$. 
Actually, from Tables~\ref{tablegra0} and \ref{tablegr} it is clear that $g_{\rho}$ is larger in the $a_0$ model than in the no-$a_0$ model for a fixed $m_0$ and $L_0$. 
In addition, $g_{\rho}$ is larger as the Yukawa coupling of $a_0$ meson $|g_{a_0 NN}|$ increases as shown in Fig.~\ref{grga0}.
Since the $\rho$ contribution $S_{\rho}$ depends on $n_B$ as shown in Eq.~(\ref{Srho}), the symmetry energy in the $a_0$ model 
increases with increasing density more rapidly than in the no-$a_0$ model.
The difference between the $a_0$ model and the no-$a_0$ model becomes transparent for smaller $m_0$, since the Yukawa coupling of $a_0$ is larger for smaller $m_0$.

\begin{figure*}
    \subfloat[$L_0$ = 40 MeV\label{Sl40}]{%
      \includegraphics[width=0.45\textwidth]{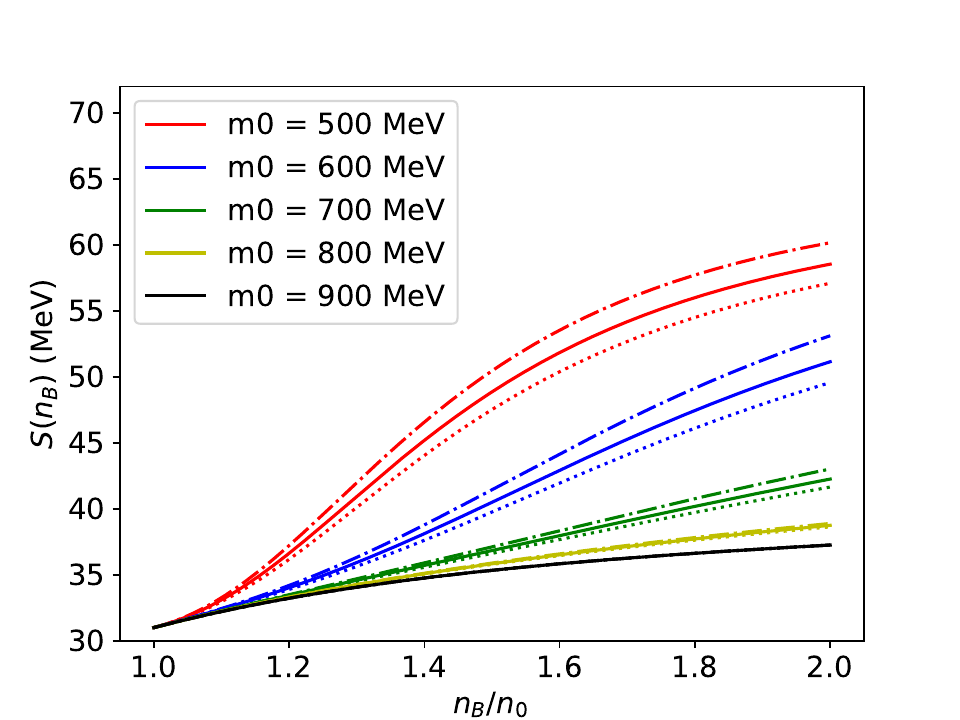}
    }
    \hfill
    \subfloat[$L_0$ = 50 MeV\label{Sl50}]{%
      \includegraphics[width=0.45\textwidth]{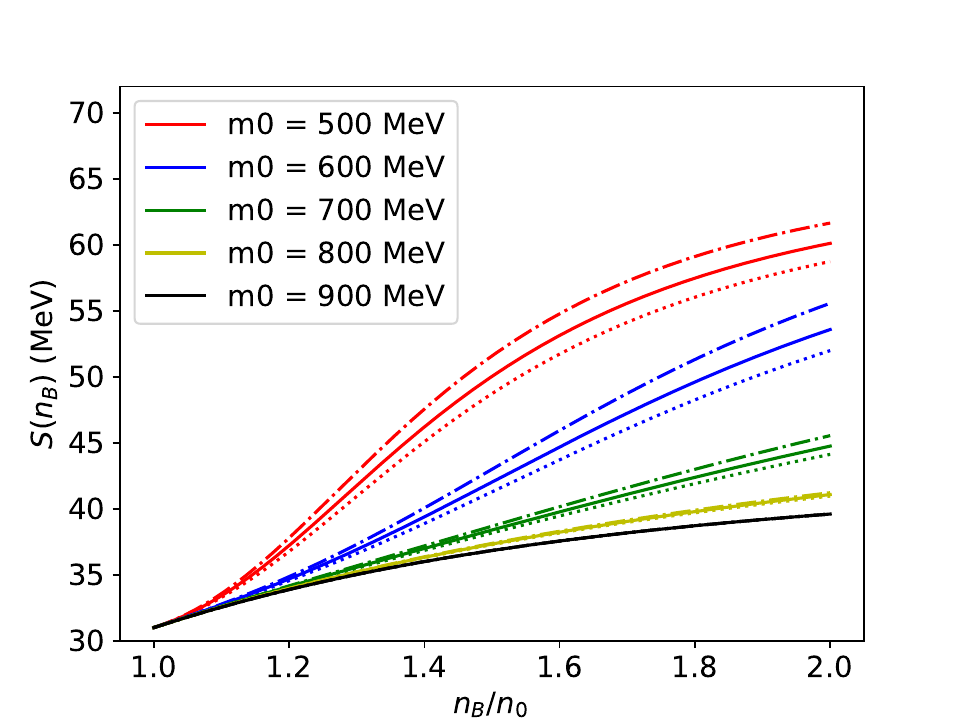}
    }
    \hfill
    \subfloat[$L_0$ = 57.7 MeV\label{Sl60}]{%
      \includegraphics[width=0.45\textwidth]{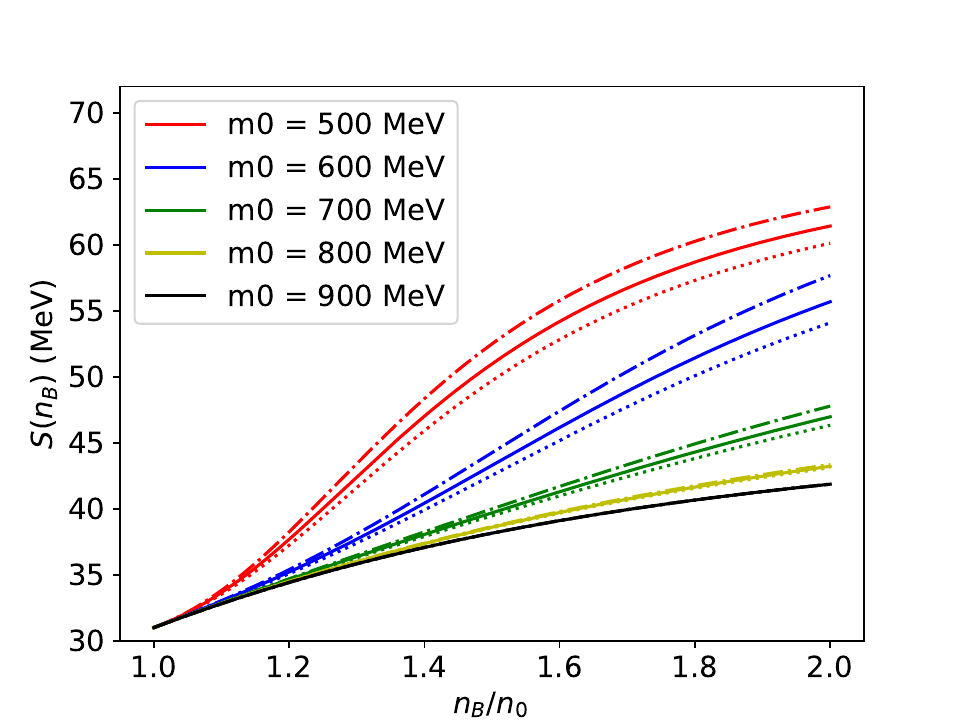}
    }
    \hfill
    \subfloat[$L_0$ = 70 MeV\label{Sl70}]{%
      \includegraphics[width=0.45\textwidth]{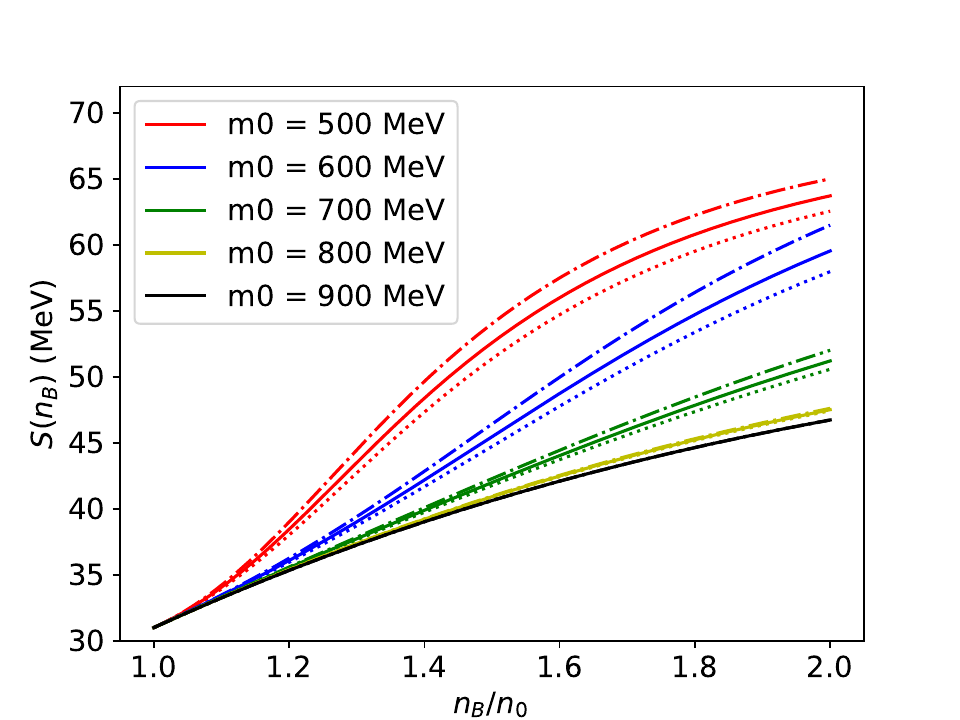}
    }
    \hfill
    \subfloat[$L_0$ = 80 MeV\label{Sl80}]{%
      \includegraphics[width=0.45\textwidth]{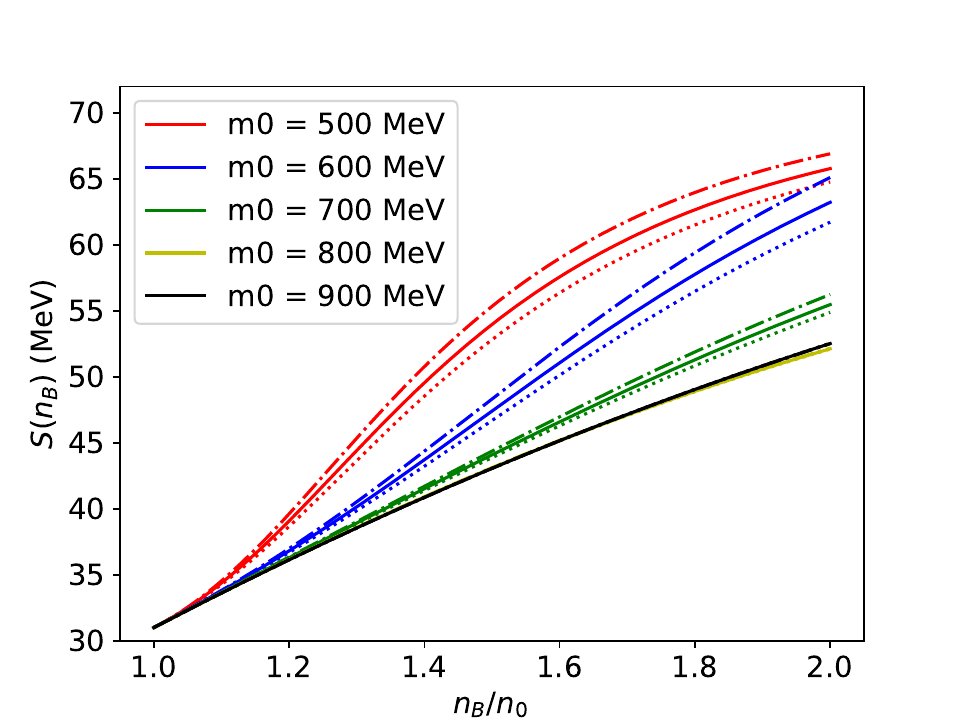}
    }
\caption{\label{symmetryenergylau6p}Density dependence of the symmetry energy $S(n_B)$ of the model including $a_0(980)$ meson for several choices of $m_0$ with $L_0= 40$-$80$\,MeV. Solid, dash-dotted, and dotted curves show the $S(n_B)$  with $\lambda_{6}'=0,\lambda_{6}$, and $-\lambda_{6}$, respectively.}
\end{figure*}

In addition, we investigate the effect of higher-order terms in the large $N_c$ expansion for the six-point interaction on the symmetry energy
by taking $\lambda_6' = \pm \lambda_6$.
The results of the symmetry energies with different values of $\lambda_{6}'$ are shown in Fig.~\ref{symmetryenergylau6p}. 
We can see that the difference between the symmetry energies for models with the same $m_0$ and $L_0$ is small, which indicates that the effect of $\lambda_6'$ on the symmetry energy is small.
Notice also that the difference becomes smaller for larger $m_0$, due to a smaller $a_0$(980) effect.

\begin{figure}[h!]
\includegraphics[scale=0.55]{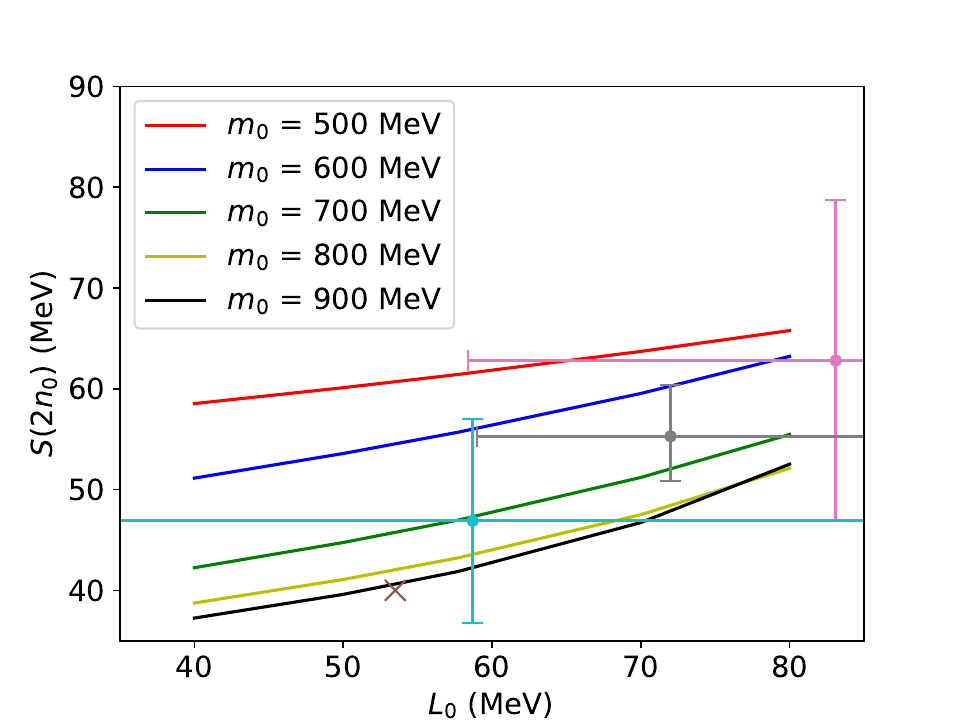}
\caption{\label{S2L0c} Comparison of $S(2n_0)$ with other models: 
Ref.~\cite{Li_2022} (brown-crosses), Ref.~\cite{PhysRevResearch.4.L022054} (pink), Ref.~\cite{PhysRevC.94.034608} (grey) and Ref.~\cite{Zhang:2019wa} (cyan).
}
\label{S2L0}
\end{figure}

At the end of this section, we compare our results of $S(2n_0)$ to the values obtained in other models in Fig.~\ref{S2L0}: 
Walecka-type RMF model with $a_0$ meson \cite{Li_2022}, 
extended Skyrme-Hartree-Fock model fitted with the PREX-II data \cite{PhysRevResearch.4.L022054}, UrQMD transport model fitted with the ASY-EOS Collaborations data \cite{PhysRevC.94.034608}, and direct inversions of observed NS radii, tidal deformability, and maximum mass in the high-density EOS space \cite{Zhang:2019wa}.
This figure shows that the results from most parameter region are consistent with those from other models.
We expect that future experimental information for $S(2n_0)$ will 
further constrain the chiral invariant mass $m_0$.

\section{effect of $a_0$(980) to the neutron star properties}
\label{section:5}

In this section, we investigate how the neutron star properties are affected by the existence of the $a_0$(980) meson. 
It is expected that $a_0$(980) meson changes the neutron star properties, as a neutron star is highly asymmetric and thus should suffer net effect from iso-spin asymmetry. 
In the following, we obtain an EoS of neutron star matter (NSM) using the present PDM, which we use in the low-density region $n_B \le 2 n_0$, and construct a unified EoS using the interpolation method adopted in Ref.~\cite{PhysRevC.103.045205,Minamikawa:2023eky}.
Then, we study the the effect of $a_0(980)$ to the $M$-$R$ relation.

\subsection{Neutron star matter EoS in the PDM}

\begin{figure}[h!]
\includegraphics[scale=0.55]{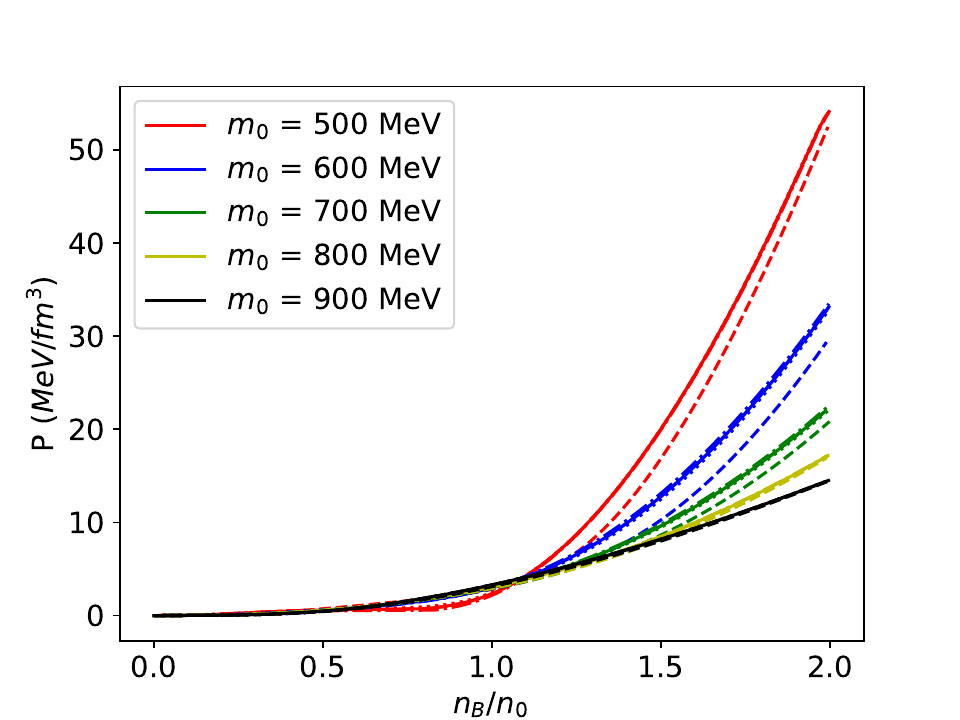}
\caption{\label{PNSMc}Comparison of pressure of the NSM, $P_{NSM}(n_B)$ with $a_0$(980) (solid curves) and without $a_0(980)$ (dashed curves) for several values of $m_0$ with $L_0=57.7$\,MeV. Dash-dotted and dotted curves show the EoS with $\lambda_{6}'=\lambda_{6}$ and $-\lambda_{6}$, respectively.}
\label{PNSM}
\end{figure}

In the low-density region $n_B \le 2 n_0$, we use the present PDM to construct the EoS for NSM.
In this work, we consider a electrically-neutral cold neutron star in equilibrium. The hadronic thermodynamic potential $\Omega_{\rm NSM}$ is given by 
\begin{equation}
\Omega_{\rm NSM}  \equiv \Omega_H + \sum_{l=e,\mu} \Omega_{l}\ ,
\label{eq38}
\end{equation} where we include the leptonic thermodynamic potentials $\Omega_{l}$ to account for electrons and muons. 
Notice that the mean fields are constrained by the stationary conditions \begin{equation}
\frac{\partial \Omega}{\partial \phi_i} \ \Bigr|_{\substack{\phi_{i}        }}  = 0 \ ,
\label{eq08}
\end{equation} where $\phi_i = (\sigma, a, \omega, \rho)$ for the matter with $a_0$(980), while $\phi_i = (\sigma, \omega, \rho)$ for matter without $a_0$(980). 
Together with charge neutrality, we obtain $\Omega_{\rm NSM}$ and thus the pressure for the neutron star: 
\begin{equation}
P_{\rm NSM} = - \Omega_{\rm NSM} \ .
\label{eq38}
\end{equation} 
The EoSs constructed with $a_0$(980) meson are shown together with the EoSs without $a_0$(980) meson in Fig.~\ref{PNSM}.
We see that the EoSs for a fixed $n_B$
are stiffened after the inclusion of $a_0(980)$ meson.
This is understood as follows:
The inclusion of $a_0(980)$ strengthens the repulsive $\rho$ meson effect to maintain the saturation properties, as mentioned in the previous section.
The attractive $a_0$ effect decreases with increasing density,
while 
the repulsive $\rho$ meson effect increases.
As a result, the pressure at a fixed $n_B$ is larger when the $a_0$(980) meson is included.

In 
Fig.~\ref{PNSM}, we also show the EoSs with $\lambda_6^{'}= \pm \lambda_6$. 
This shows that the effect of $\lambda_6^{'}$ to the EoS is small in most of the cases. This is expected because the high-order interactions are suppressed in the larger $N_c$ limit.

\subsection{Unification of neutron star EoS}
\label{section:5b}

Following Refs.~\cite{PhysRevC.103.045205,Minamikawa:2023eky}, we interpolate the EoS obtained from the present model in the low-density region $n_B \leq 2 n_0$ and the one from the SU(3)$_L\times$SU(3)$_R$ NJL-type quark model for $n_B \geq 5 n_0$, to construct the unified EoS 
exhibiting hadron-quark continuity. We think that it is not suitable to use the PDM in higher density region since hyperons, which are not included in the present PDM, may appear at such high density region.
Here, we briefly summarize the way of interpolation.
(See, for detail,  Refs.~\cite{PhysRevC.103.045205,Minamikawa:2023eky}.)

In the interpolating regime, the pressure $P_I$ is expanded as a function of $\mu_B$ as \begin{equation}
    P_{I}(\mu_B) \equiv \sum^5_{i=0} c_i \mu_B^i \ ,
\end{equation} 
where $c_i$ are coefficients to be determined. This $P_I(\mu_B)$ is connected to the hadronic EoS $P_H(\mu_B)$ at $\mu_B=\mu_B^L$, where $\mu_B^L$ is the chemical potential corresponding to the density twice of the normal nuclear density, $n_B = 2n_0$, by requiring
\begin{equation}
\begin{aligned}
P_I(\mu_B=\mu_B^L) &= P_H(\mu_B=\mu_B^L)\ ,\\
\frac{\partial^n P_{I}}{\partial \mu_B^n} \Bigr|_{\substack{\mu_B=\mu_B^L}} &= \frac{\partial^n P_{H}}{\partial \mu_B^n} \Bigr|_{\substack{\mu_B=\mu_B^L}} \ , \quad (n=1,2) \ .
\end{aligned}
\end{equation} 
Similarly, $P_I(\mu_B)$  is connected to $P_{Q}(\mu_B)$ constructed from the NJL-type quark model by requiring
\begin{equation}
    \frac{\partial^n P_{I}}{\partial \mu_B^n} \Bigr|_{\substack{\mu_B=\mu_B^H}} = \frac{\partial^n P_{Q}}{\partial \mu_B^n} \Bigr|_{\substack{\mu_B=\mu_B^H}} \ , \quad (n =0,1,2) \ ,
\end{equation} 
where $\mu_B^H$ is the chemical potential corresponding to $n_B = 5n_0$. 
We accept the EoS as physical one when the sound velocity $c_s$ calculated as
\begin{equation}
    c_s^2 \equiv \frac{\partial P}{\partial \epsilon} = \frac{\partial P}{\partial \mu_B}\frac{\partial \mu_B}{\partial n_B}\frac{\partial n_B}{\partial \epsilon} = \frac{1}{\mu_B}\frac{n_B}{\frac{\partial^2 P}{\partial \mu_B^2}}\ , 
\end{equation} 

\begin{figure*}
    \subfloat[$m_0$ = 500 MeV\label{H500}]{%
      \includegraphics[width=0.45\textwidth]{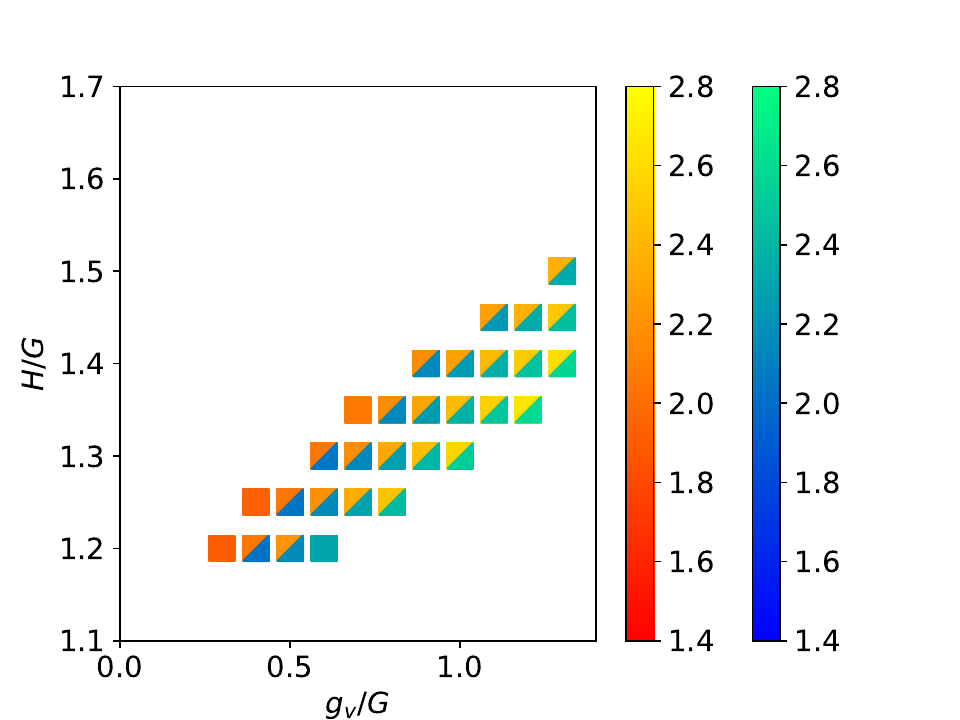}
    }
    \hfill
    \subfloat[$m_0$ = 600 MeV\label{H600}]{%
      \includegraphics[width=0.45\textwidth]{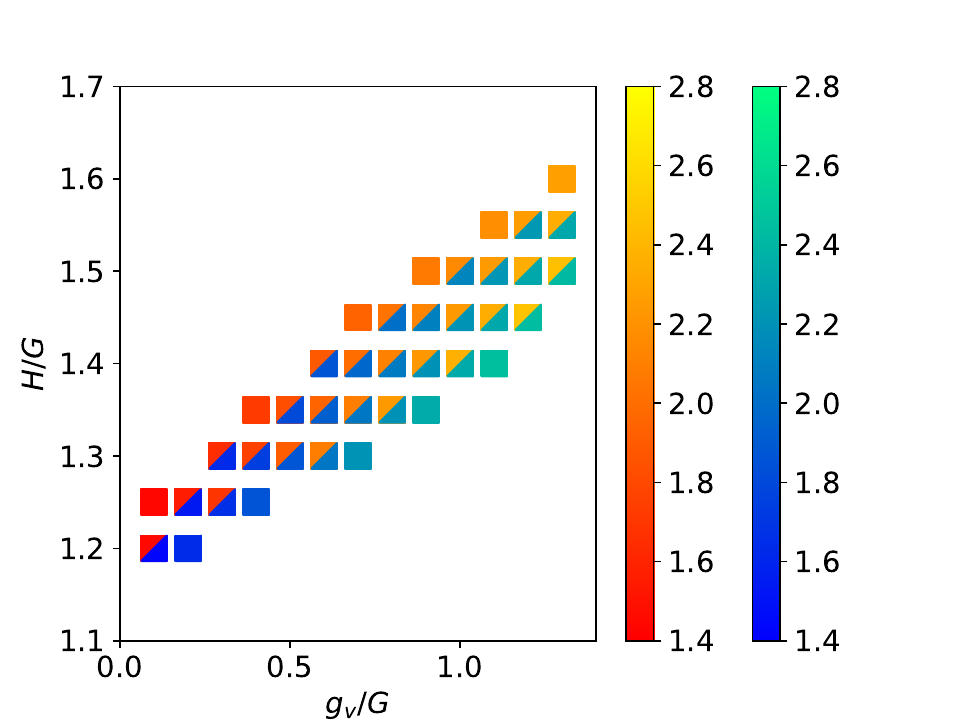}
    }
    \hfill
    \subfloat[$m_0$ = 700 MeV\label{H700}]{%
      \includegraphics[width=0.45\textwidth]{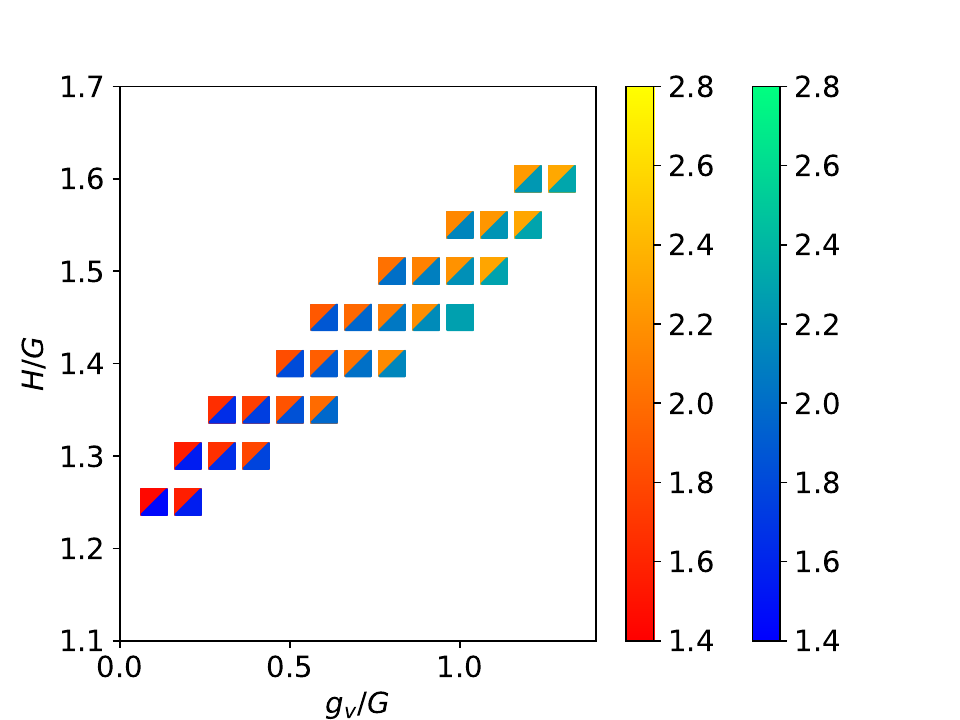}
    }
    \hfill
    \subfloat[$m_0$ = 800 MeV\label{H800}]{%
      \includegraphics[width=0.45\textwidth]{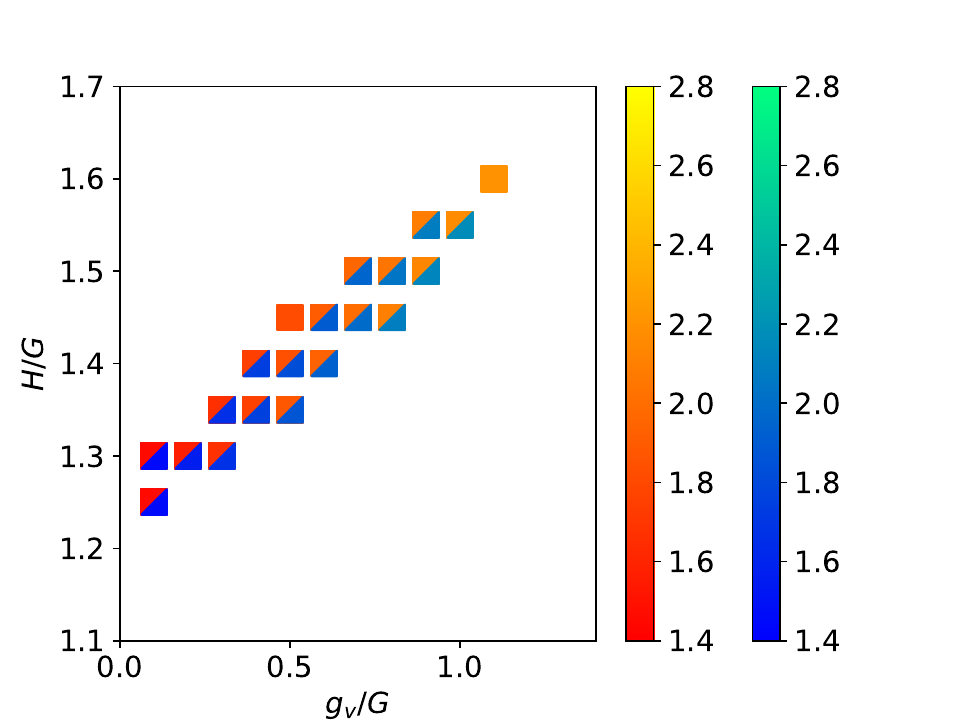}
    }
    \hfill
    \subfloat[$m_0$ = 900 MeV\label{H900}]{%
      \includegraphics[width=0.45\textwidth]{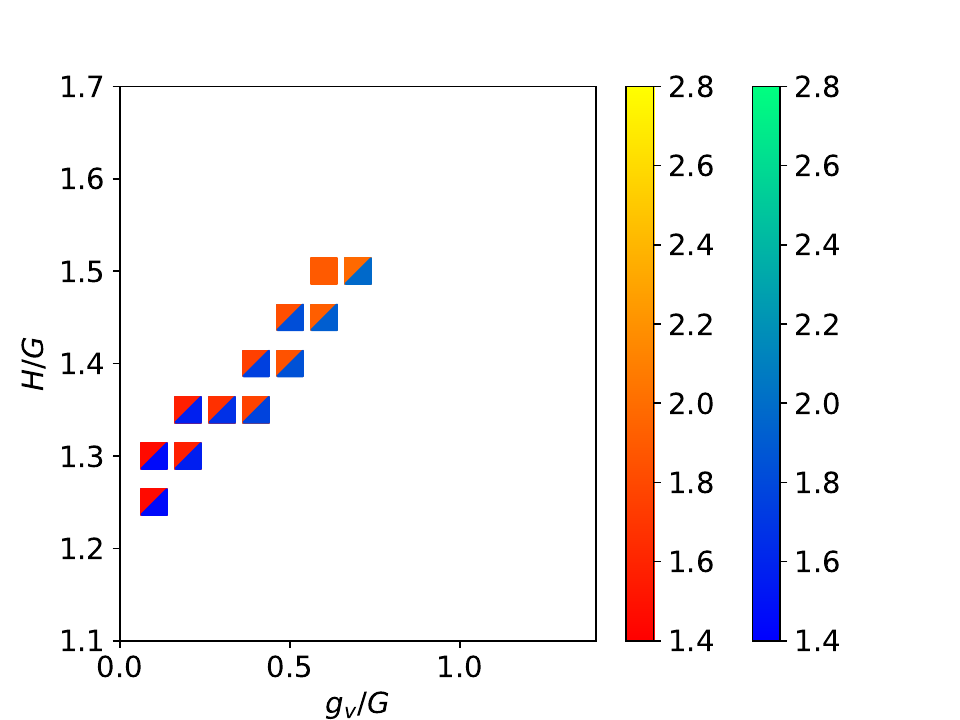}
    }
\caption{\label{Hgvc}
(color online) Allowed range of NJL parameters ($H/G$,$g_v/G$) for several choices of $m_0$ with $L_0=57.7$ MeV, $\lambda_6'=0$. \includegraphics[width=0.016\textwidth]{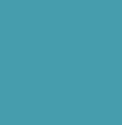} shows the allowed parameters for the model with $a_0(980)$, while \includegraphics[width=0.016\textwidth]{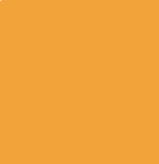} shows the allowed parameters for the model without $a_0(980)$.  \includegraphics[width=0.0165\textwidth]{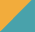} represents the parameters that are allowed for both of the models. 
The color indicates the maximum mass of the corresponding neutron star in the unit of solar mass. 
}\label{Hgv}
\end{figure*}

is 
smaller than the speed of light:
\begin{equation}
    0 \leqslant c_s \leqslant 1 \ .
\end{equation} 
%

This requirement restrict the range of NJL parameters $(H/G,g_v/G)$, which are shown in Fig.~\ref{Hgv}. In the figure, the winter-colored squares and autumn-colored squares show the allowed parameters for the model with and without $a_0(980)$, respectively. This shows a positive correlation between $H/G$ and $g_v/G$, which agrees with previous studies \cite{Baym_2019, PhysRevC.103.045205, PhysRevC.106.065205, Minamikawa:2023eky}. 
We also notice that the allowed NJL parameter sets of $a_0$ model favor slightly larger $g_v$ compare to the model without $a_0(980)$ meson because the hadronic EoSs are slightly stiffened as shown in Fig.~\ref{PNSM}.



\begin{figure}[h!]
\includegraphics[scale=0.55]{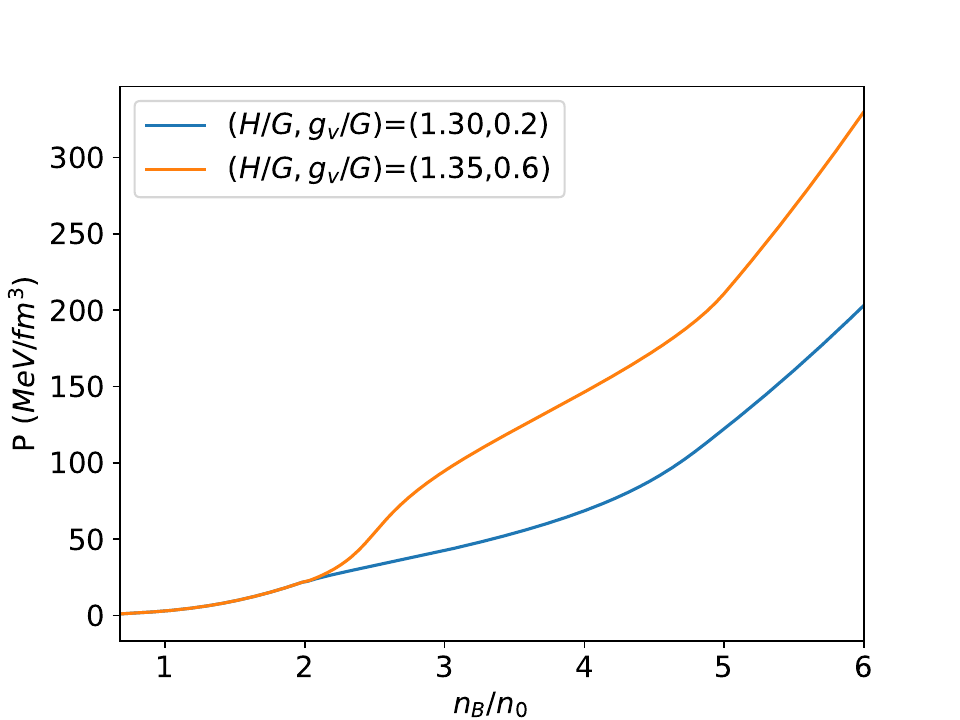}
\caption{\label{fig:wide}Examples of interpolated pressure of neutron star matter with $a_0$(980) for $m_0=700$\,MeV, $L_0=57.7$\,MeV and $\lambda_{6}'=0$.
}
\label{PNS}
\end{figure}
\begin{figure}[h!]
\includegraphics[scale=0.55]{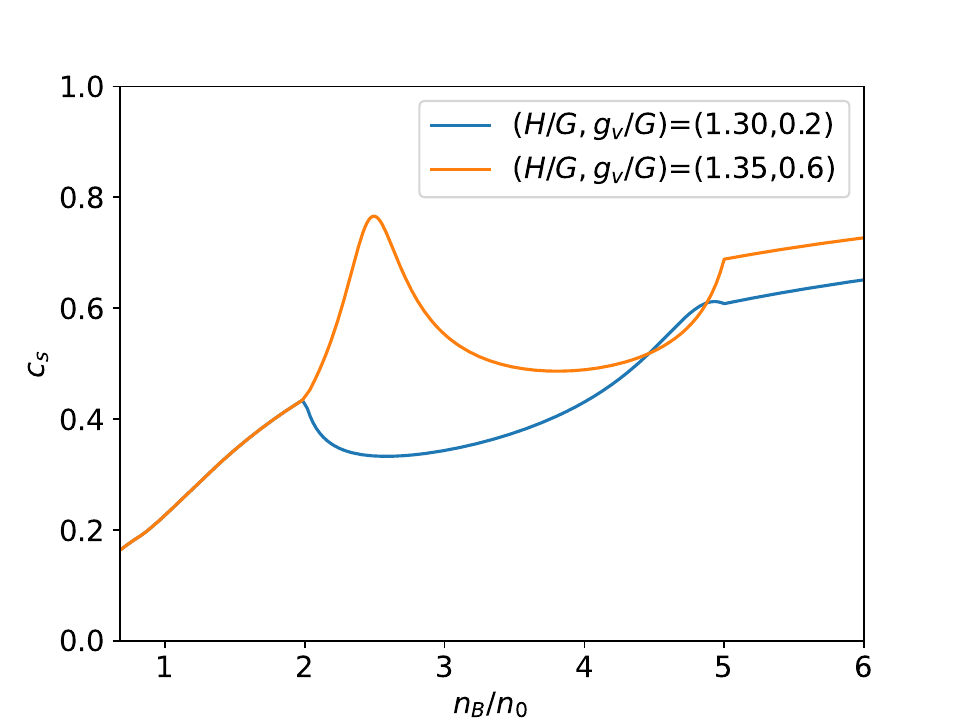}
\caption{\label{fig:wide}Speed of sound $c_s$ corresponding to the interpolated pressure in Fig.~\ref{PNS}.}
\label{CS}
\end{figure}

Finally in this subsection, we show some typical examples of interpolated EoSs in Fig.~\ref{PNS} with the corresponding speed of sound in Fig.~\ref{CS}.

\subsection{Mass-Radius relation}

In this subsection, we compute the $M$-$R$ relation by solving the Tolman-Oppenheimer-Volkoff (TOV) equation~\cite{PhysRev.55.364,PhysRev.55.374}, and compare the results with the observational data.

\begin{figure}[h]
\includegraphics[scale=0.55]{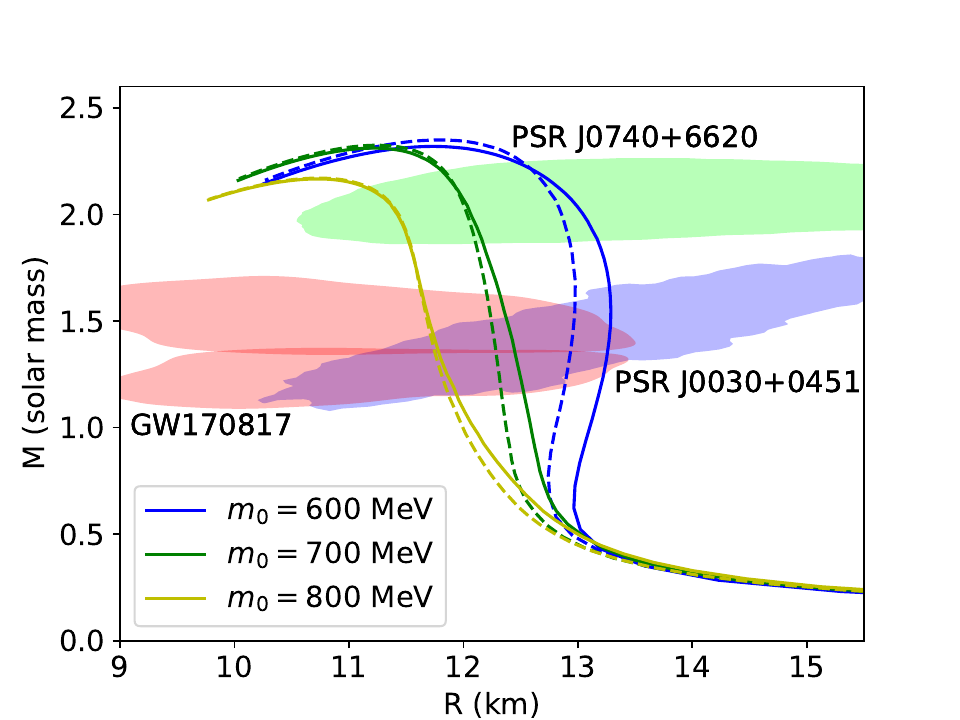}
\caption{\label{30c}$M$-$R$ relations computed from models with and without the existence of $a_0(980)$ for different $m_0$ with $L_0=57.7$ MeV. Solid curves represent  the $M$-$R$ relations from the model with $a_0(980)$ meson and dashed curves the ones of the model without $a_0(980)$. Parameters ($H/G$,$g_v/G$) of the NJL-type quark model are chosen as to be the same for a specific $m_0$, $(1.55,1.3)$ for $m_0 = 600$\,MeV, $(1.6,1.3)$ for $m_0 = 700$\,MeV, and $(1.55,1)$ for $m_0 = 800$\,MeV, respectively.
}
\label{30}
\end{figure}

We first examine how the existence of $a_0(980)$ affects to the $M$-$R$ relation. 
In 
Fig.~\ref{30}, we show some typical examples of the $M$-$R$ relations computed with and without the existence of $a_0(980)$ for $m_0=600$, $700$ and $800 $\,MeV where $\lambda'_6=0$ and $L_0 = 57.7$\,MeV are taken.
%
This figure clearly shows that 
inclusion of the $a_0(980)$ meson 
has increased 
the radius for the neutron stars with the mass of $0.5 \lesssim M/M_{\odot} \lesssim 2$,  
by the amount of $\lesssim 1$\,km depending on the parameters. 
We see that the difference becomes smaller for larger $m_0$ similarly to the one for the symmetry energy.
This is because the $a_0(980)$ meson couples to the matter weaker for large $m_0$. 
In addition, we observe that the $a_0$(980) meson has little effect on the maximum mass, since the core of such heavy neutron star includes the quark matter and the maximum mass is mainly determined by the parameters of the NJL-type quark model.

\begin{figure}[h]
\includegraphics[scale=0.55]{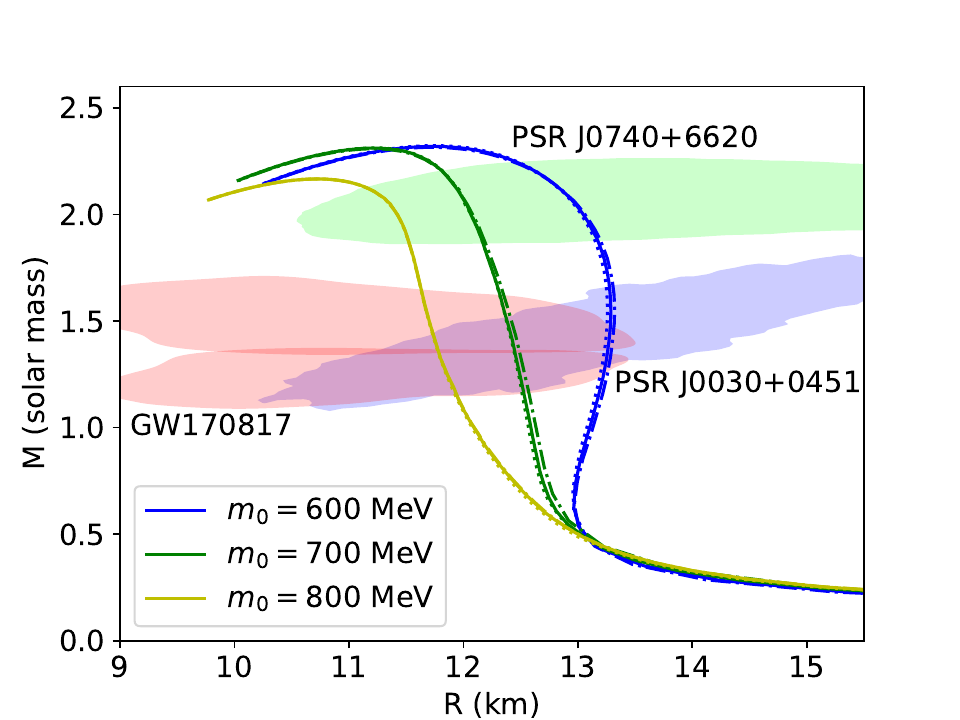}
\caption{\label{31c}$M$-$R$ relations computed from models including $a_0$(980) with different choices of $\lambda'_6$. 
Solid curves represent the results for $\lambda_{6}'=0$, dashed-dottted curves for $\lambda_{6}'=+\lambda_6$ and dotted curves for $\lambda_{6}'=-\lambda_{6}$. 
NJL parameters ($H/G$,$g_v/G$) are chosen as $(1.55,1.3)$ for $m_0 = 600$\,MeV, $(1.6,1.3)$ for $m_0 = 700$\,MeV, and $(1.55,1)$ for $m_0 = 800 $\,MeV. The slope parameter is fixed as $L_0 = 57.7$\,MeV.
}
\label{31}
\end{figure}

We also 
study the effect of high-order interaction in the large $N_c$ limit to the $M$-$R$ relation.
Figure~\ref{31} shows the computed $M$-$R$ relations of the models with different choices of $\lambda'_6$.
As expected, we find that $\lambda'_6$ has small effect on the $M$-$R$ relation and changes the radius by only $\lesssim 0.5$\,km.

We compare the results to the observational constraints on neutron star mass and radius obtained from PSR J0030+0451 by \cite{Miller_2019}, PSR J0740+6620 by \cite{Miller_2021}, and GW170817 by \cite{Abbott_2019}, which restricts the value of the chiral invariant mass as 
\begin{equation}
   580 \,\text{ MeV} \lesssim m_0 \lesssim 860 \,\text{ MeV} \ . 
\end{equation} 
for $L_0=57.7$\,MeV and $\lambda'_6=0$. The constraint is increased by $\sim 200$ MeV comparing to the results of refs.~\cite{PhysRevC.106.065205,Minamikawa:2023eky} considering the $U(1)_A$ anomaly. The constraints for different choices of $L_0$ are summarized in Table~\ref{tablem0a0}.
\begin{table}[h!]
\caption{\label{tablem0a0}Summary of constraints of the chiral invariant mass $m_0$ for the models with different $L_0$. The unit of $m_0$ is in MeV.}
\begin{ruledtabular}
\begin{tabular}{cccc}
&\textrm{With $a_0$}
&\textrm{Without $a_0$}\\
\colrule
$L_0$ = 40 MeV &  $510  \lesssim m_0 \lesssim 830$ &  $ 500 \lesssim m_0 \lesssim 840$ \\
$L_0$ = 50 MeV& $560  \lesssim m_0 \lesssim 840 $  & $ 520 \lesssim m_0 \lesssim  850$  \\
$L_0$ = 60 MeV& $580  \lesssim m_0 \lesssim 860 $  & $540 \lesssim m_0 \lesssim 870$  \\
$L_0$ = 70 MeV& $610  \lesssim m_0 \lesssim 890 $ & $ 570 \lesssim m_0 \lesssim 910$  \\
$L_0$ = 80 MeV& $640  \lesssim m_0 \lesssim 940 $ & $610 \lesssim m_0 \lesssim 950$ \\
\end{tabular}
\end{ruledtabular}
\end{table}
This shows
that the effect of $a_0$ meson increases the lower bound by about $10$-$40$\,MeV and reduces the upper bound by about $10$-$20$\,MeV. We note that, for $L_0 = 80$\,MeV in the no-$a_0$ model, the saturation properties cannot be satisfied when $m_0 \gtrsim 960$\,MeV.

In the present study, we used the interpolation density for PDM as $n^{PDM}_I = 2n_0$ to avoid the direct consideration of hyperons.
Here, we study the dependence of our results on the choice of $n^{PDM}_I$ by taking $n^{PDM}_I = 1.5n_0$ and $2.5n_0$.
We find that, when  the interpolation density of PDM is increased, the allowed values of $(H/G,g_v/G)$ are narrowed down. Accordingly, decreasing the interpolating density from $n_I^{PDM} = 2n_0$ to 1.5 $n_0$ reduces the lower limit of $m_0$ constraint of $a_0$ model with $L_0=$ 57.7 MeV by 20 MeV and increases the upper limit by 90 MeV, while increasing $n_I^{PDM}$ from $2n_0$ to 2.5 $n_0$ does not change the lower limit of allowed $m_0$ and reduces the upper limit by 50 MeV. We think that it is not suitable to use a high interpolation density since the details of hyperon appearance in neutron star is not well-known.
Then the above results show that the dependence of the lower limit of $m_0$ is not so sensitive to the choice of the interpolating density.
On the other hand, although the upper limit has some dependence, the stability of the nuclear matter at normal nuclear density requires $m_0 \lesssim 950$\,MeV.
%
Thus, the constraint to the chiral invariant mass is summarized as
$m_0 \gtrsim 500$ MeV for small $L_0$, while $m_0 \gtrsim 600$ MeV for large $L_0$.

\section{Summary and discussions}\label{section:6}

We constructed a PDM with $a_0$ meson based on the chiral SU(2)$_L\times$SU(2)$_R$ symmetry, and studied the effect of $a_0$(980) on the symmetry energy and the neutron star properties. 
We showed that, for $m_0 \lesssim 800$\,MeV, the symmetry energy in the density region $n_B \le 2 n_0$ is increased by the inclusion of  $a_0(980)$ meson, which leads to the stiffening of the EoS of neutron star matter resulting an enlargement of the radius of neutron star by about $1$\,km. 
This stiffening is understood as follows:
The $a_0$ meson provides an attractive force to reduce the symmetry energy, which is balanced with the repulsive force by the $\rho$ meson at the saturation density. Then, the repulsive force  by the $\rho$ meson is stronger in the $a_0$ model than in the no-$a_0$ model.
Since the $a_0$ contribution decreases with increasing density while the $\rho$ contribution increases, the symmetry energy increases more rapidly in the model with $a_0$ than the model without $a_0$.
As a result, the symmetry energy is larger when the $a_0$ meson is included.
Furthermore, since the Yukawa coupling of the $a_0$ meson to the nucleon is larger for smaller chiral invariant mass $m_0$, the stiffening effect of $a_0$ meson is stronger for smaller $m_0$.
%
%
We compared the resultant $M$-$R$ relation to the neutron star observational data from PSR J0030+0451, PSR J0740+6620, and the gravitational wave event GW170817.

Including ambiguity of the interpolating density, below which the EoS is obtained from the present PDM, we conclude that   
the constraint to the chiral invariant mass is summarized as
$m_0 \gtrsim 500$ MeV for small $L_0$, while $m_0 \gtrsim 600$ MeV for large $L_0$.

We also found 
that, for 
$m_0 = 900$\,MeV, the symmetry energy is slightly reduced by including the effect of $a_0$ meson, due to the softening effect of $\omega$-$\rho$ mixing. This implies that the behavior of the symmetry energy for $n_B > n_0$ depends on the subtle balance among the effects of the $a_0$ meson attraction and  the $\rho$ meson repulsion combined with the $\omega$-$\rho$ mixing.
%
We note that the softening effect by the $a_0$ meson is also reported in Refs.~\cite{Li_2022,https://doi.org/10.48550/arxiv.2209.02861}.

In the present analysis, we have included three terms for the 
six-point scalar meson interaction.
As we showed, the effects from 
$tr[(M^\dagger M )^2]tr[M^\dagger M ]$ and $\{ tr[M^\dagger M ] \}^3$, which are suppressed in the large $N_c$ limit, 
are indeed small to the symmetry energy and the neutron star EoS. Therefore, these high-order effect may ignored in the future to simplify the model.



\begin{acknowledgments}

We would like to thank Ryota Yoda who performed primary analysis in his master thesis. 
This work was supported in part by JSPS KAKENHI Grant No. 20K03927. T.M. was also supported by JST SPRING, Grant No. JPMJSP2125.

\end{acknowledgments}

\bibliography{PDM-a0-refs}

\end{document}